\documentclass[]{aa}
\usepackage{graphicx}
\usepackage{txfonts}
\usepackage{lipsum}
\usepackage{subcaption}         
\usepackage{lscape}             
\usepackage{placeins}           
\usepackage{mwe}
\usepackage{hyperref}
\usepackage{url}
\usepackage{xcolor}
\hypersetup{
colorlinks,
linkcolor={red!80!black},
citecolor={blue!80!black},
urlcolor={blue!80!black}
}

\def\cm3{cm$^{-3}$}
\def\kms{km~s$^{-1}$}

\def\msun{M$_{\odot}$}

\def\one{\ts {\,\sc i}}
\def\two{\ts {\,\sc ii}}
\def\three{\ts {\,\sc iii}}
\def\four{\ts {\,\sc iv}}
\def\five{\ts {\sc v}}

\def\beq{\begin{equation*}}
\def\eeq{\end{equation*}}

\def\lesssim{\mathrel{\hbox{\rlap{\hbox{\lower4pt\hbox{$\sim$}}}\hbox{$<$}}}}
\def\gtrsim{\mathrel{\hbox{\rlap{\hbox{\lower4pt\hbox{$\sim$}}}\hbox{$>$}}}}
\def\mic{\,$\mu$m}

\def\one{{\,\sc i}}
\def\two{{\,\sc ii}}
\def\three{{\,\sc iii}}
\def\four{{\,\sc iv}}
\def\five{{\sc v}}

\def\ergs{erg\,s$^{-1}$}

\newcommand{\iso}[2]{\ensuremath{^{#1}\rm{#2}}}

\def\apjs{ApJS}
\def\apjl{ApJL}
\def\aap{A\&A}

\def\nifs{\iso{56}Ni}

\def\lya{Ly\,$\alpha$}
\def\ha{H$\alpha$}
\def\pag{Pa\,$\gamma$}

\def\heinir{He\one\,$\lambda$\,$10830.2$}

\def\niidoub{[N\two]\,$\lambda\lambda$\,$6548.0,\,6583.5$}
\def\niiauroral{[N\two]\,$\lambda$\,$5755$}

\def\oidoub{[O\one]\,$\lambda\lambda$\,$6300.3,\,6363.8$}

\def\oiidoub{[O\two]\,$\lambda\lambda$\,$7319.5,\,7330.2$}

\def\oiiidoub{[O\three]\,$\lambda\lambda$\,$4958.9,\,5006.8$}

\def\naid{Na\one\,$\lambda\lambda\,5896,5890$}

\def\neiiiopt{[Ne\three]\,$\lambda$\,$3868.8$}

\def\mgiiuv{Mg\two\,$\lambda\lambda$\,$2795,\,2802$}

\def\stwoopt{[S\two]\,$\lambda$\,$4068.6$}
\def\siinir{[S\two]\,$\lambda\lambda\,10286.7,\,10370.5$}
\def\siiinir{[S\three]\,$\lambda$\,$9530.62$}

\def\caiidoub{[Ca\two]\,$\lambda\lambda$\,$7291.5,\,7323.9$}

\def\feiiopt{[Fe\two]\,$\lambda$\,$7155.2$}

\def\nkiiopt{[Ni\two]\,$\lambda\lambda$\,$7377.8,\,7411.6$}

\def\mhenv{$M$(H-env)}
\def\mcds{$M_{\rm CDS}$}
\def\vcds{$V_{\rm CDS}$}
\def\mej{$M_{\rm ej}$}
\def\ekin{$E_{\rm kin}$}

\begin{document}

\title{The diversity of late-time interaction in Type II supernovae}

\author{
    Luc Dessart\inst{\ref{inst1},\ref{inst1a}}\corrauth{dessart@iap.fr}
\and
Wynn V. Jacobson-Gal\'an\inst{\ref{inst2},\ref{inst3}}
\and
K. Azalee Bostroem\inst{\ref{inst4},\ref{inst5}}
}

\institute{
    French-Chilean Laboratory for Astronomy, IRL 3386, CNRS and Instituto de Astrofísica,
       Pontificia Universidad Católica de Chile, Casilla 306, Santiago, Chile.\label{inst1}
\and
Institut d'Astrophysique de Paris, CNRS-Sorbonne Universit\'e, 98 bis boulevard Arago, F-75014 Paris, France\label{inst1a}
  Cahill Center for Astrophysics, California Institute of Technology, Pasadena, CA 91125, USA\label{inst2}
  \and
  NASA Hubble Fellow\label{inst3}
\and
Steward Observatory, University of Arizona, 933 North Cherry Avenue, Tucson, AZ 85721-0065, USA\label{inst4}
\and
LSST-DA Catalyst Fellow\label{inst5}
}

   \date{Received; accepted}

  \abstract{
Studying core-collapse supernovae (SNe) beyond one year after explosion, when ejecta interaction with circumstellar material (CSM) progressively becomes the dominant power source, offers critical insights into the  progenitor mass loss history. Using binary progenitors that underwent  different levels of H-rich envelope stripping, we present radiative transfer calculations of the ejecta with CSM interaction in  Type IIb, II-L and II-P SN scenarios covering epochs of 1.5 to 3--6\,yr, with a focus on the UV and optical properties. The impact of CSM interaction is evaluated for a cold-dense shell (CDS) of 0.1--2.0\,\msun\ (CDS velocity of $\sim$\,$5000-10\,000$\,\kms) and shock powers of $10^{39}-10^{41}$\,\ergs, producing a grid of about 200 models. We find that \lya, \mgiiuv, and \ha\ are typically the strongest individual lines, \mgiiuv\ being the most constant relative to changes in time and in CDS mass. \lya\ thrives at low CDS mass and is strongly attenuated otherwise. A large fraction of the emerging UV and optical flux is, however, influenced by a forest of Fe lines, which introduces strong optical depth effects. A myriad of other lines, generally forbidden, are predicted that each contain less than a percent of the total flux. N enrichment yield stronger N lines in the UV and optical. A subset of these models roughly match UV and optical observations of SNe 1993J, 1998S, 2013ej, and 2023ixf at 1.5--3\,yr, although matching the full range consistently is a challenge, in particular when dust must be invoked. Blue-red asymmetry of line profiles is typical of all lines in the UV, the more so at earlier times or higher CDS mass, even in the absence of dust. In our simulations, optical depth effects, associated with the gas or dust or both, quenches the emission of \lya\ or \oiiidoub\ from regions within or below the CDS.
}

\keywords{
  supernovae: general --
  supernovae: individual: SN\,1993J, SN\,1998S, SN\,2013ej, SN\,2023ixf --
  Radiative transfer --
  Dust --
  Line: profiles
}
   \maketitle

\nolinenumbers


\section{Introduction}
\label{sect_intro}

Most of the observations of core-collapse supernovae (SNe) are focused on their early phase  of evolution when they are optically bright (see, for example, \citealt{anderson_2pl}; \citealt{wynn_pap2_24}; \citealt{hinds_csm_25}; \citealt{das_2p_25}). Because of the exponential decline of radioactive decay power, SN ejecta become increasingly faint and eventually undetectable on timescales of months to years. SN\,1987A is an exception in this respect as it continues to be observed today only because of its proximity (see, e.g., \citealt{abellan_87A_17}) -- it would be undetectable at any representative SN distance of 10\,Mpc or more. Consequently, SNe may be detected at later times only if there is another power source, either from the compact remnant (e.g., the Type Ib SN\,2012au; \citealt{milisavljevic_12au_18}) or from interaction with circumstellar material (CSM; e.g., the Type II-P SN\,2017eaw; \citealt{weil_17eaw_20}). Such old SNe can thus shed light on the physics of young compact objects or on the progenitor mass loss history. Furthermore, injected power is probe of the material reprocessing that power (e.g. progenitor composition), providing different constraints from early time observations. There is thus much interest in studying SNe at late times, on year timescales or more, as they transition to a young supernova remnant.

For the current sample of late-time detections in core-collapse SNe, and nearly exclusively in Type II SNe (there are rare cases such as the Type Ib SN\,2014C; \citealt{milisavljevic_14C_15,margutti_14C_16}), the power source seems to be interaction with CSM.
In events initially classified as Type IIn SNe, the objects continue to be detected until late times suggesting the presence of a large reservoir of CSM out to large distances. One such example is SN\,1998S which was interacting from first light \citep{leonard_98S_00,fassia_98S_01} and for years afterwards \citep{fransson_cno_93J_98S_05,smith_ptf11iqb_15}.
In some other events, the signatures of interaction with CSM are evident at first light (i.e., presence of narrow lines with electron-scattering broadened wings) but ebb after a few days. The transient then evolves as what appears to be a noninteracting SN for months until clear signs of interaction emerge again after 1--2 years. One such event is SN\,2023ixf, which showed IIn signatures for about a week (see, e.g., \citealt{jacobson_galan_23ixf_23,bostroem_23ixf_23}), followed by a photospheric evolution in which the signatures of interaction were limited to a slight excess in the UV and some excess flux in \ha\ \citep{bostroem_23ixf_24}. And it is only later on in the nebular phase that the UV flux came to dominate the spectral energy distribution (i.e., the CSM interaction came to dominate the radioactive decay), with a morphing of line profiles in the optical from narrow to broad and boxy \citep{kumar_23ixf_25,folatelli_23ixf_25,wynn_sed_25,zheng_23ixf_25,bostroem_uv_26}. Other events show no signs of interaction until about two years. However, the optical brightness then switches from an exponential decline to a plateau while the spectra exhibit a morphological change from narrow emission lines powered by radioactive decay in the inner ejecta to broad and boxy emission profiles arising from the outer ejecta (e.g., SN\,2004et, \citealt{kotak_04et_09}; SN\,2013ej, \citealt{mauerhan_13ej_17}; SN\,2017eaw, \citealt{weil_17eaw_20}).

The picture described above is, however, strongly sensitive to the wavelength range where the observations are undertaken. The UV is thought to dominate the escaping radiation of year-old SNe with interaction \citep{chevalier_fransson_94,dessart_csm_22,dessart_late_23}, and thus optical observations alone may not be conclusive. Furthermore, dust is thought to form both in the inner, metal-rich ejecta as well as in the outer ejecta (or the cold dense shell -- CDS -- that forms at the interface with the CSM), causing both absorption in the optical and thermal emission in the IR (e.g., \citealt{lucy_dust_89}; \citealt{sarangi_dust_22}; \citealt{sarangi_cds_22}; \citealt{shahbandeh_jwst_23}).

Numerical modeling of the radiation arising from Type II SNe interacting with CSM on a 1--10\,yr timescale have been carried out since the 1990s. \citet{chevalier_fransson_94} modeled the evolution of UV-optical-IR line fluxes for the first decades following the explosion of a standard red-supergiant star and the interaction of the ejecta with the progenitor wind. \citet{chevalier_82_radio_xray} studied the radio and X-ray emission properties of Type II SNe (for a review see \citealt{chevalier_fransson_rev_17}), with specific studies of nearby events like SN\,1993J \citep{fransson_93j_96}. \citet{fransson_cno_93J_98S_05} presented UV and optical observations of SNe 1993J and 1998S extending out to a few years and found evidence for CNO processing from UV emission line ratios. But most of these studies focused on the properties of the high-energy radiation that escapes, and much less on the UV and optical properties of the associated transients. In particular, none of these works provide a complete modeling  of the spectra of the low-energy radiation, from UV to optical and infrared.

Our approach differs in that it leaves aside the details of the power injection and focuses instead on the reprocessing of the shock power by the ejecta and CDS and how this shapes the SN spectrum from the UV to the IR. The method was presented in \citet{dessart_csm_22} and \citet{dessart_late_23} and recently applied to SN\,2023ixf \citep{dessart_23ixf_phot_26,dessart_23ixf_neb_26}. One motivation for using this approach is that late-time SNe II with interaction exhibit mostly emission lines in the UV and optical from material of modest ionization, suggesting that the shock power at the origin of the UV and optical radiation is efficiently thermalized. Consequently, this approach makes no prediction for the escaping X-ray radiation.

The majority of massive stars are born in binaries that will interact during their evolution (e.g., see \citealt{sana_bin_12,sana_bin_25}), leading to a transfer of mass to the companion (or mass loss by the binary system) that far exceeds the impact of wind mass loss in stars of moderate mass, or to the merger of the two stars (e.g., \citealt{zapartas_bin_19}). Thus, to extend our previous work on a generic 15\,\msun\ single star that exploded as a massive red supergiant \citep{dessart_csm_22,dessart_late_23}, we here employed progenitor models that evolved in a binary system and reached core collapse with different levels of stripping, yielding final H-rich envelope masses that span from about 0.1 to 7\,\msun\ \citep{ercolino_bin_24}. In the absence of any CSM, these exploding stars produce a range of SN photometric and spectroscopic properties that mimic the trend from SNe IIb, to II-L and II-P \citep{dessart_bin_24} -- see also, for example, \citet{eldridge_sn2_18} or \citet{hiramatsu_2p_2l_21}. In this work, we explore how these different SNe evolve from 1.5 to 3--6 years under the influence of interaction with CSM, varying for that purpose the CDS mass and velocity, or the power injected. We thus document the broad diversity of UV and optical signatures of ejecta interaction with CSM for the full range of Type II SNe and their associated supergiant-star progenitors.

This paper is structured as follows. In Section~\ref{sect_setup} we present the numerical setup. Section~\ref{sect_res_time} describes the spectral evolution from 1.5 year (i.e., 500\,d) to about 6 years for one representative model. In Section~\ref{sect_res_mcds}, we use one of our models to explore the impact of varying the CDS mass at a given epoch. In Section~\ref{sect_res_logP}, we explore how these results differ when the mass of the progenitor H-rich envelope is varied. The subsequent sections present specific comparisons to contemporaneous UV and optical observations of SN\,1993J (Section~\ref{sect_res_93J}), SN\,1998S (Section~\ref{sect_res_98S}), SN\,2013ej (Section~\ref{sect_res_13ej}; optical data only), and SN\,2023ixf (Section~\ref{sect_res_23ixf}). In Section~\ref{sect_conc}, we present our conclusions. Additional illustrations are provided in the appendix in Section~\ref{sect_more}.


\begin{figure}
\includegraphics[width=0.49\textwidth]{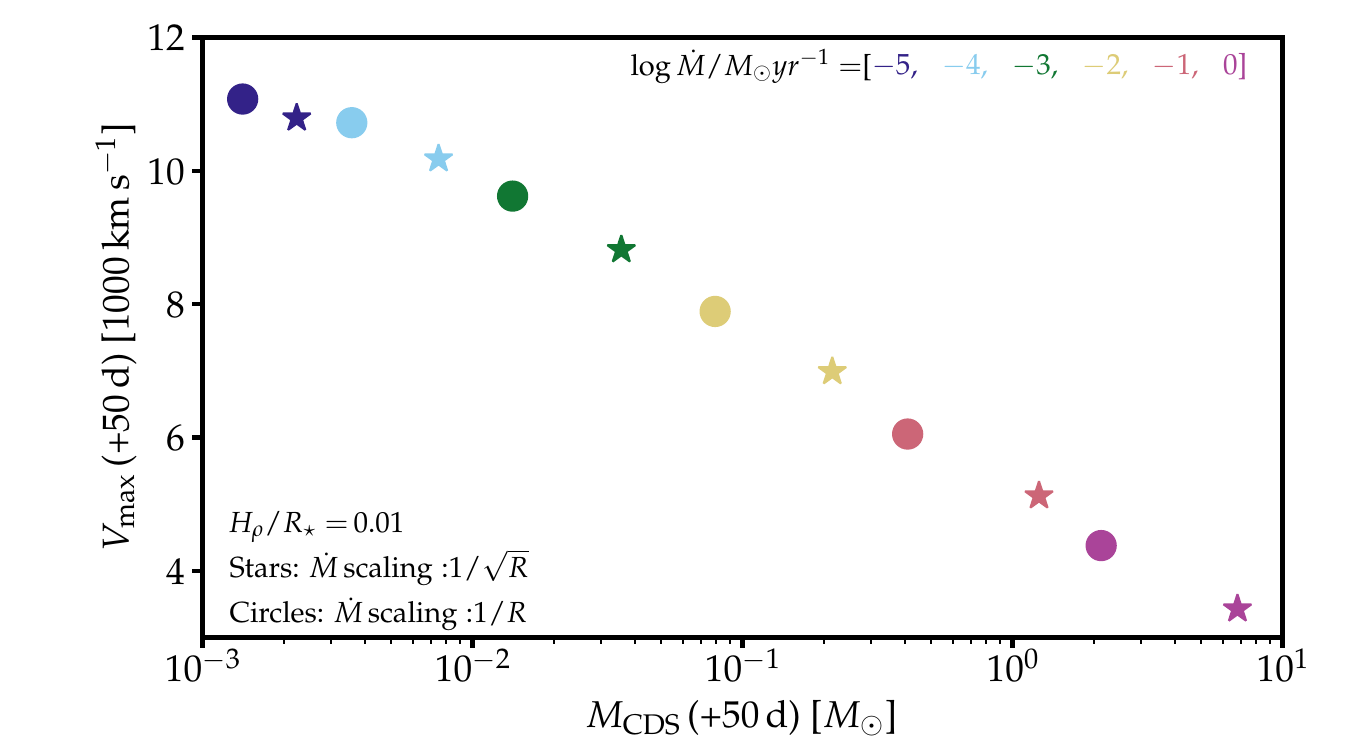}
\caption{Illustration of the maximum ejecta velocity versus the CDS mass at 50\,d in the simulations of a standard red-supergiant star enshrouded within various amounts of CSM \citep{dessart_wynn_23}. The labels indicate the adopted wind mass loss rate values, the additional CSM density scaling of $1/r$ or $1/\sqrt{r}$ that was used, as well as the density scale height at the stellar surface.
\label{fig_hydro}
}
\end{figure}

\section{Numerical setup}
\label{sect_setup}

The numerical approach used for the present radiative-transfer simulations is identical to that presented in \citet{dessart_23ixf_phot_26,dessart_23ixf_neb_26} for the modeling of SN\,2023ixf at photospheric and nebular epochs, and given in more details in \citet{dessart_csm_22} and \citet{dessart_late_23}. The treatment of shock power injection is either in the form of high-energy electrons deposited with a prescribed spatial distribution (as in \citealt{dessart_csm_22,dessart_late_23}) or in the form of X-rays (for details, see Appendix~A of \citealt{dessart_pm_24}, as well as Section~\,4 of \citealt{dessart_23ixf_neb_26}) -- the former is used by default and the latter invoked to cure specific discrepancies of the former (e.g., the presence of \lya\ at early times in the nebular phase of SN\,2023ixf \citep{dessart_23ixf_neb_26} or the presence of \oiiidoub) or to gauge the uncertainties associated with the treatment of power injection. In this work, the spatial distribution of the X-ray power injection has a gaussian profile, which by default was centered on the CDS and with a standard deviation of 35\,\kms. This centroid was sometimes shifted away from the CDS (typically by a few 10\,\kms) to bias the power injection in favor of the reverse shock or of the forward shock. Such modulations were carried out in Section~\ref{sect_obs} to test the sensitivity of our results and exploring possible ways of improving the agreement with observations. We thus constrain the nature, magnitude, and distribution of the injected power by confronting to UV and optical observations.

A fundamental difference with our previous work on late-time SNe II interacting with CSM is that we now broaden the range of envelope stripping of the progenitors. For a given explosion energy, this leads to a different ejecta structure and composition, including the mean expansion rate, the velocity of the H-rich material, or the He and N abundance in the H-rich material. Observationally, it covers from SNe of Type IIb to those of Type II-P and thus the broadest possible range in H-rich envelope mass for Type II SNe. Specifically, we used the models of \citet{dessart_bin_24}, which originate from a subset of binary massive star evolution models presented in \citet{ercolino_bin_24}. Our sample is differentiated in terms of initial binary period, which in days and in the log have the value of 2.85, 2.90, 2.95, 3.0, 3.1, 3.2, 3.35, and 3.45 (these models are thus named logP2p85 etc).

\begin{table*}
  \caption{Set of simulations performed in this study.
\label{tab_set}
}
\begin{center}
\begin{tabular}{|l|c|c|c|c|c|c|c|}
\hline
Progenitor & $M_{\rm f}$ & \mhenv\   & $X_{\rm s}$(He) & \vcds\        & \mcds\  & Epochs  &   Power  \\
\hline
           & [\msun]  & [\msun]  &                 & [\kms]        & [\msun] &  [d]    & [\ergs]  \\
\hline
logP2p85   & 3.92  & 0.19   &   0.48          &   10643       &     0.2    &   550 $\rightarrow$ 976    & 1e40, 3e39 \\
logP2p90   & 4.06  & 0.28   &   0.44          &   10492       &     0.2    &   550 $\rightarrow$ 976    &    1e40, 3e39 \\
logP2p95   & 4.14  & 0.31   &   0.41          &   10476       &     0.2    &   550 $\rightarrow$ 1181   &    1e40, 3e39 \\
logP3p00   & 4.26  & 0.39   &   0.39          &    9834       &     0.2    &   550 $\rightarrow$ 1074   &    1e40, 1e39 \\
logP3p10   & 4.51  & 0.60   &   0.36          &    9670       &     0.2    &   550 $\rightarrow$ 1299   &    1e40, 3e39 \\
logP3p20   & 5.18  & 1.18   &   0.32          &    9231       &     0.2    &   550 $\rightarrow$ 1299   &    1e40, 3e39, 5e38 \\
logP3p35   & 7.97  & 3.74   &   0.30          &    6966       &     0.2    &   550 $\rightarrow$ 1429   &    1e40, 3e39 \\
logP3p45   & 10.84 & 6.86   &   0.30          &    6390       &     0.2    &   550 $\rightarrow$ 2000   &    1e40, 3e39 \\
\hline
logP3p35   & 7.97  & 3.74   &   0.30          &    7627       &     0.1    &   550 $\rightarrow$ 887    &    1e40  \\
logP3p35   & 7.97  & 3.74   &   0.30          &    6966       &     0.2    &   550 $\rightarrow$ 1429   &    1e40, 3e39  \\
logP3p35   & 7.97  & 3.74   &   0.30          &    6325       &     0.4    &   550 $\rightarrow$ 1902   &    1e40, 3e39  \\
logP3p35   & 7.97  & 3.74   &   0.30          &    5984       &     0.6    &   550 $\rightarrow$ 1074   &    1e40, 3e39  \\
logP3p35   & 7.97  & 3.74   &   0.30          &    5743       &     0.8    &   550 $\rightarrow$ 1729   &    1e40, 3e39  \\
logP3p35   & 7.97  & 3.74   &   0.30          &    5563       &     1.0    &   550 $\rightarrow$ 1902   &    1e40, 3e39  \\
logP3p35   & 7.97  & 3.74   &   0.30          &    5222       &     1.5    &   550 $\rightarrow$ 1729   &    1e40, 3e39  \\
logP3p35   & 7.97  & 3.74   &   0.30          &    5002       &     2.0    &   550 $\rightarrow$ 1902   &    1e40, 3e39  \\
\hline
\end{tabular}
\end{center}
Notes: The table gives some details about the progenitor star, the ejecta, and the CDS parameters used in our simulations. From left to right, the columns first give the final mass, the H-rich envelope mass, and the surface He mass fraction of the progenitor stars at core collapse, followed by the CDS velocity and mass, and finally the epochs at which the radiative transfer calculations were carried out together with the range of powers adopted (either in the form of high-energy, nonthermal electrons or as X-rays). In all cases, the original ejecta (i.e., without a CDS) contains 0.08\,\msun\ of \nifs\ and have a kinetic energy of 10$^{51}$\,erg.
\end{table*}

\begin{figure}
\includegraphics[width=0.49\textwidth]{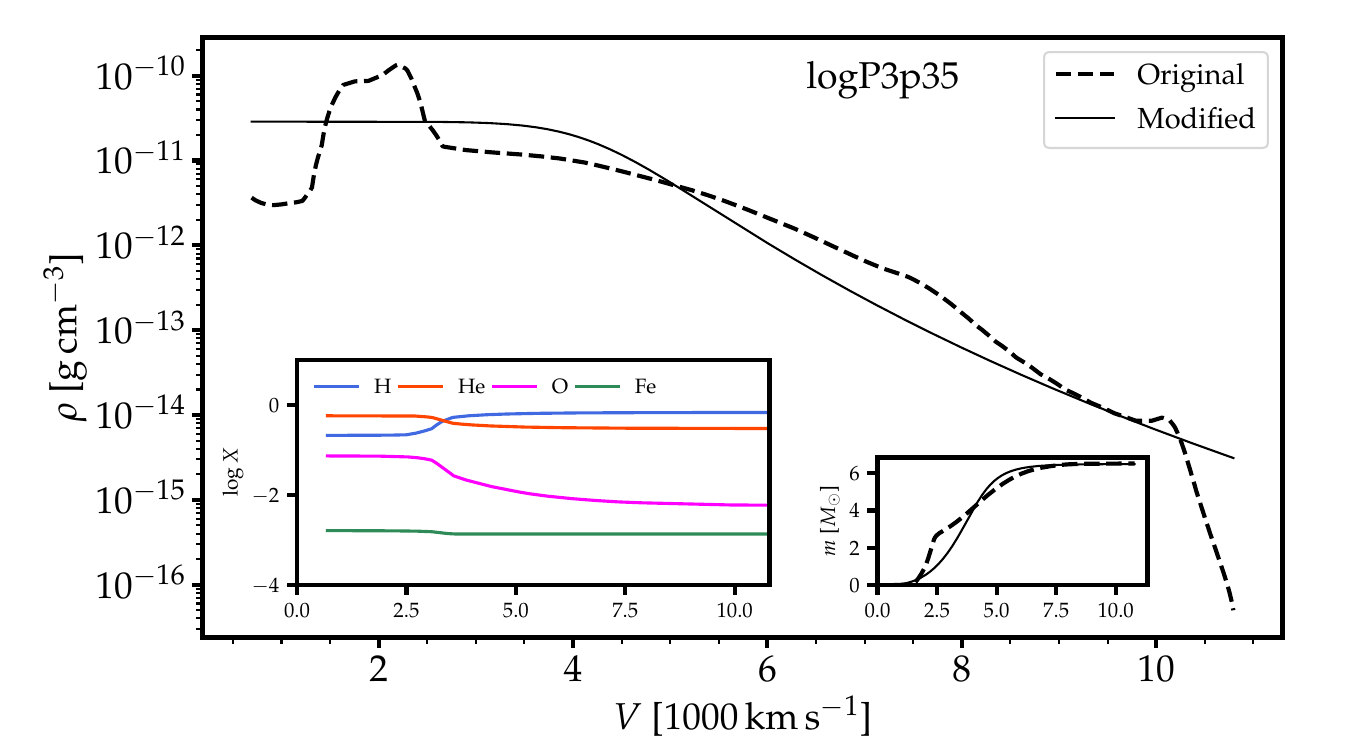}
\includegraphics[width=0.49\textwidth]{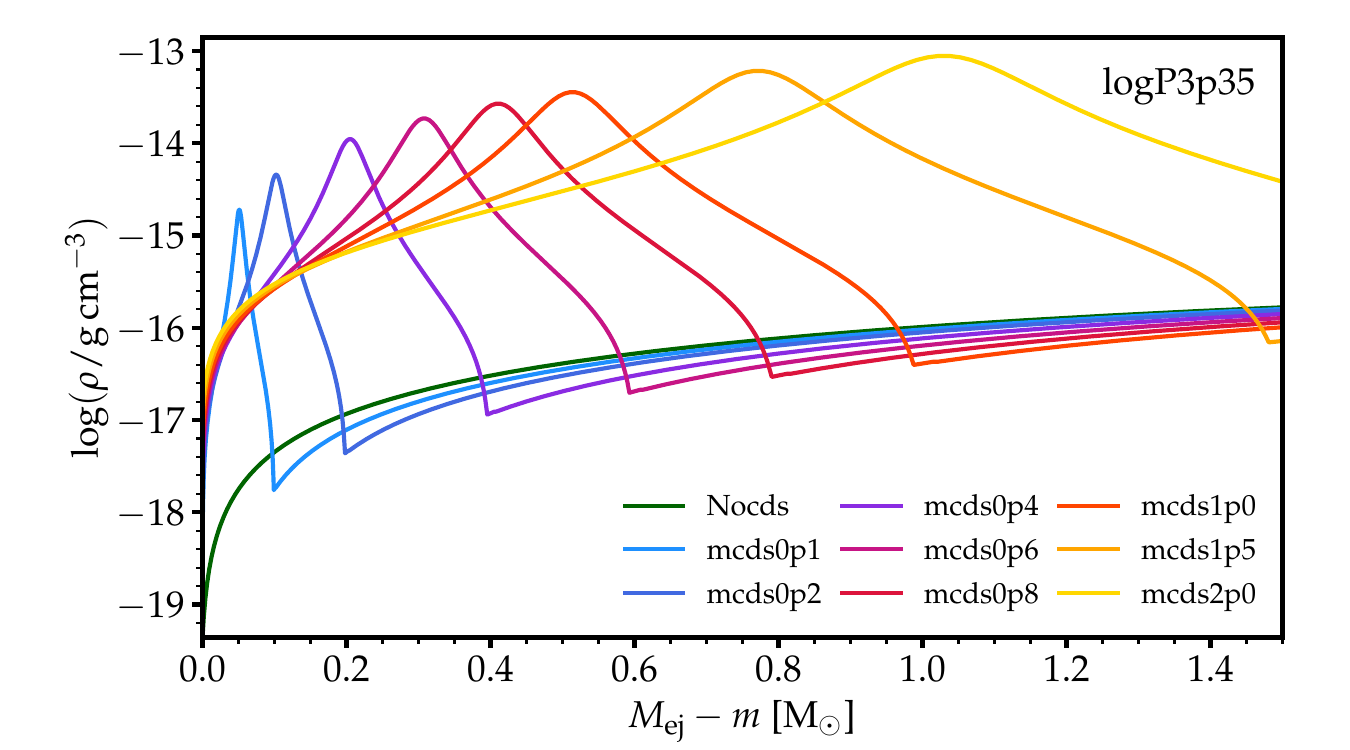}
\caption{Top: Original and modified density profile versus velocity for model logP3p35 without a dense shell and at 12.1\,d. The insets show the composition and the Lagrangian mass versus velocity. Bottom: Illustration of the density profile versus Lagrangian mass (integrated from the surface inward) for the model logP3p35 at 500\,d and endowed with a CDS mass rising from zero (``Nocds'') up to 2.0\,\msun\ (``mcds2p0''). Here and in similar figures, the term \mej\ in the $x$-axis refers to the total debris mass, thus the original ejecta plus any swept mass from the CSM.
\label{fig_prog}
}
\end{figure}
\begin{figure}
\includegraphics[width=0.49\textwidth]{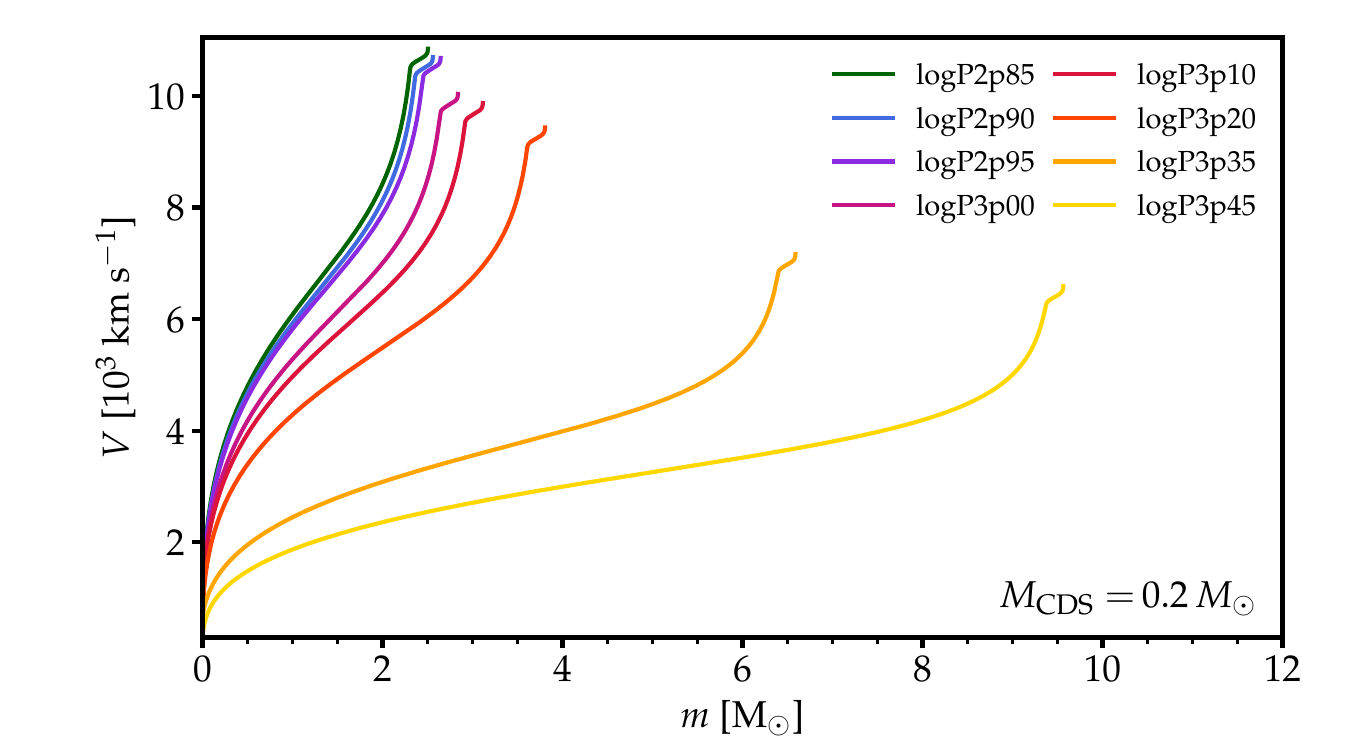}
\includegraphics[width=0.49\textwidth]{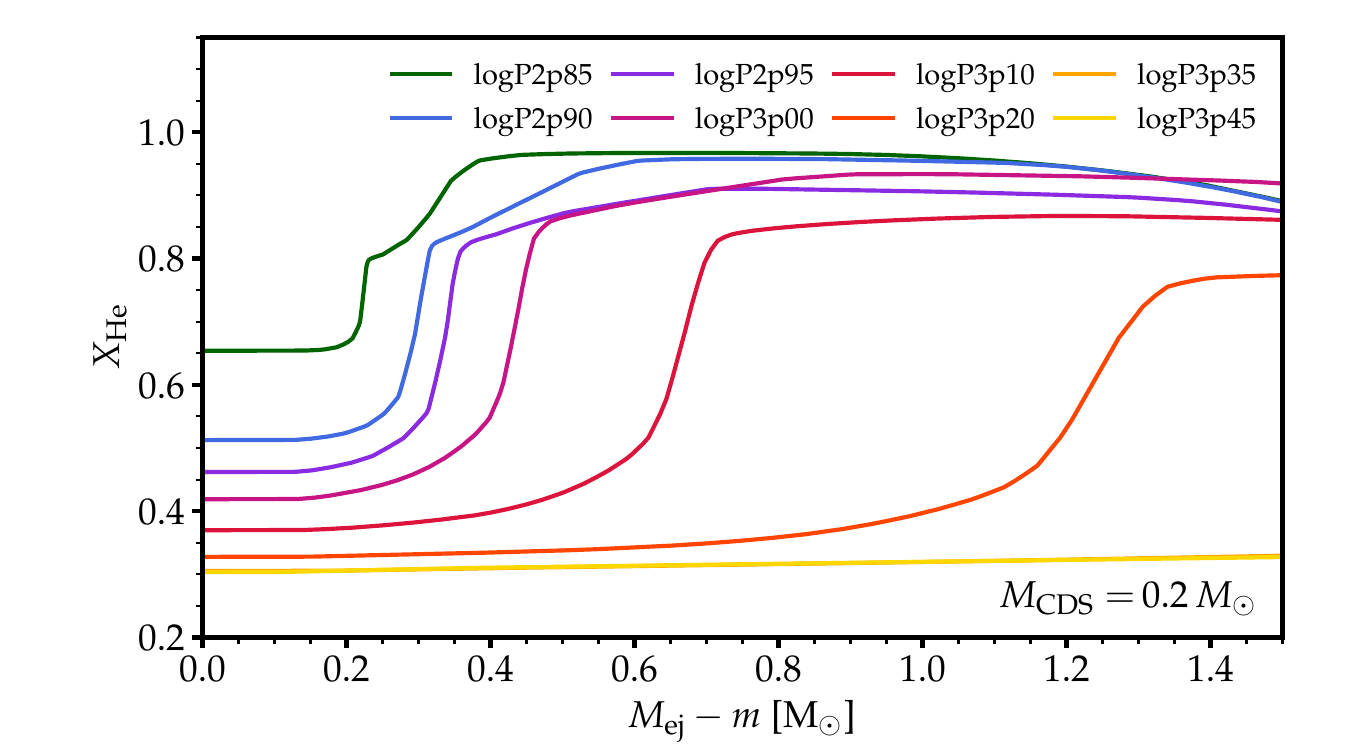}
\includegraphics[width=0.49\textwidth]{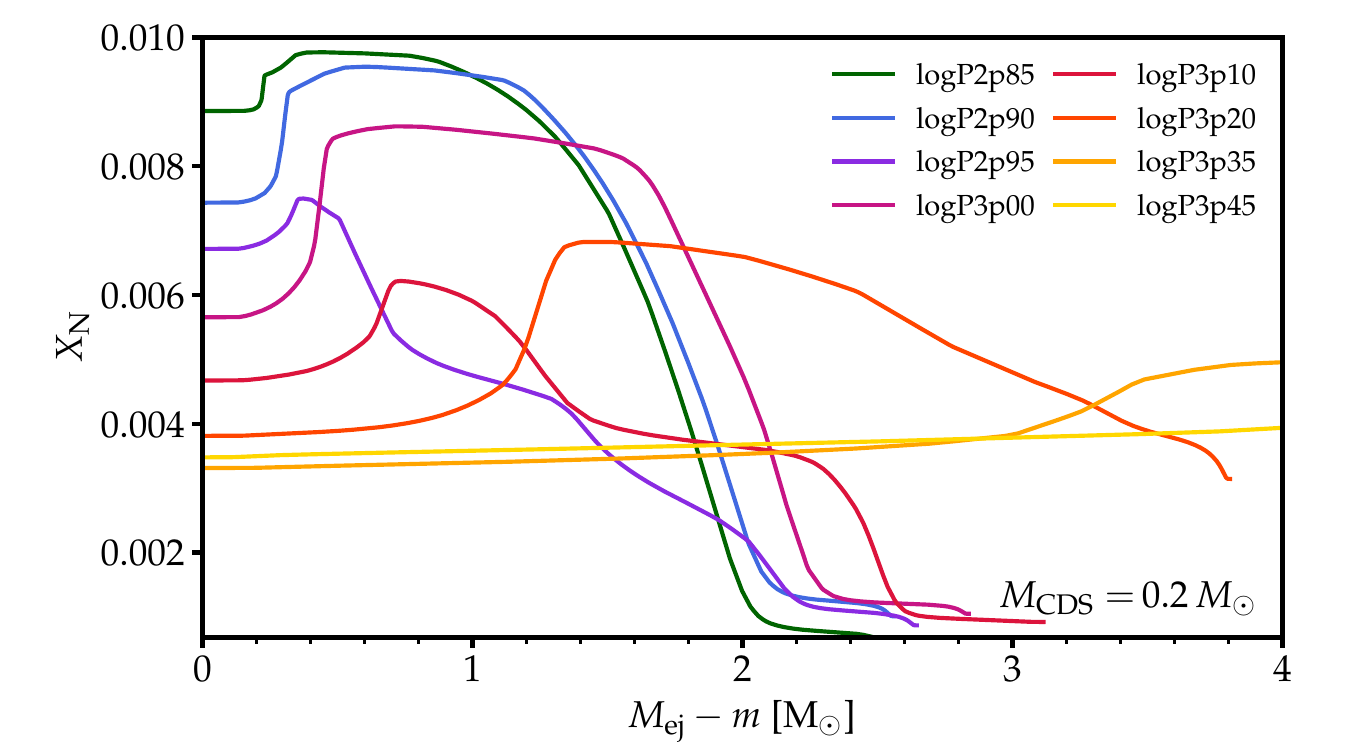}
\caption{Illustration of the velocity (top) and the He (middle) and N (bottom) abundance profiles versus Lagrangian mass (integrated inwards or outwards -- see xlabel) for our grid of progenitor models with varying \mhenv\ at core collapse (i.e., models logP2p85 to logP3p45) and augmented with a CDS of 0.2\,\msun.
\label{fig_other_prog}
}
\end{figure}

To allow for a flexible, parameterized approach, we make a number of simplifications. First, to avoid the sharp variations in density that arise from 1D radiation-hydrodynamics of red-supergiant star explosions (see Fig.~2 of \citealt{dessart_bin_24}), we reset their density structure using the procedure of \citet{chugai_hv_07}. That is, adopting the \mej\ and \ekin\ of the original model, a new ejecta density structure is calculated, and the original composition is then remapped in Lagrangian mass space onto that new structure -- the procedure is then repeated for all ejecta models in our sample. The \nifs\ abundance profile was reset to enforce strong mixing for the same mass of 0.08\,\msun, followed by a renormalization to maintain the sum of mass fractions at unity. Only the \nifs\ radioactive decay chain is considered here and associated isotopes were decayed to the start time of 500\,d used in this work for all simulations. These simplifications were motivated by the fact that the model of \citet{dessart_late_23} computed at an age of 1000\,d took one full month to converge because of the complex shuffled-shell composition structure (for details see \citealt{dessart_shuffle_20}), which requires about 350 grid points, as well as the numerical difficulties arising from the outer CDS where shock power is injected. Here, with the smooth composition and density structure of our ejecta, only 70 points are necessary (leaving aside for now the CDS) and numerical convergence is greatly facilitated, allowing us to explore a broad parameter space.

All these models were then augmented with a CDS, with mass \mcds\ and velocity \vcds. Rather than running radiation-hydrodynamics simulations from shock breakout out to late times, we approximated the structure at the time of 500\,d. Based on results from previous radiation-hydrodynamics  simulations of SNe II with CSM (see Fig.~\ref{fig_hydro} here, or the Fig.~2 of \citealt{dessart_wynn_25}), we assumed here that a given CSM mass would decelerate the same amount of ejecta material, and that the CDS would be centered at a velocity corresponding to that Lagrangian mass in the ejecta (i.e., $M_{\rm ej} - 0.5 M_{\rm CDS}$). Hence, half the CDS is composed from the CSM and the other half from the ejecta. All else being the same, a greater CDS mass corresponds to a deeper CDS location (i.e., a greater deceleration of the outer ejecta) and a closer proximity to the metal-rich ejecta regions. Obviously, this effect is exacerbated in progenitors with a lower \mhenv. In this work, the bare ejecta models of \citet{dessart_bin_24} were augmented by a CDS with a mass that covers from 0.1 up to 2.0\,\msun. Not all combinations were considered.

The CDS density profile was set to have a gaussian shape with a centroid at \vcds\ and a standard deviation of 70\,\kms. The material within the CDS was also clumped such that the volume filling factor was given by $1 - (1-a) \exp(-X^2)$, with $X=(V-V_{\rm CDS}) / dV_{\rm CDS}$, $dV_{\rm CDS}=$\,100\,\kms, and $a=0.01$ -- this ensures strong clumping within the CDS and none exterior to it. From models logP2p85 to logP3p45 (thus spanning \mhenv\ from 0.19 to 6.86\,\msun) and augmented with a 0.2\,\msun\ CDS, the velocity of the CDS decreases from 10\,643 to 6390\,\kms. For model logP3p35 augmented with an increasing CDS mass from 0.1 to 2.0\,\msun, the centroid velocity of the CDS decreases from 7627 to 5002\,\kms. Our present simulations are not meant to be perfectly hydrodynamically consistent (it is obviously possible to compute large grids of simulations for many configurations, as done for one progenitor star in \citealt{dessart_wynn_25}, but this is unnecessary for the present conceptual study -- none of our results rest, for example, on the claim that the CDS moves at $X$\,\kms\ or 5\,\% faster or slower etc) nor do they reflect consistently the progenitor mass loss history, which is never known accurately anyway.

Figure~\ref{fig_prog} illustrates the properties of model logP3p35 for a variety of CDS masses, as well as the impact of going from the 1D hydrodynamical model of \citet{dessart_bin_24} to the smooth, analytical counterpart with the same \mej, \ekin, and abundance profiles in mass space. Figure~\ref{fig_other_prog} shows how the different progenitors differ in structure and composition for the same adopted CDS mass of 0.2\,\msun. For the same adopted \ekin\ and \mcds, the CDS velocity is systematically greater for lower \mej\ (enhanced progenitor stripping). The abundances of He and N are also notably greater in the CDS for more stripped progenitors.

We typically used a constant power injection inside the CDS at all times, with a representative value of 10$^{40}$\,\ergs. Additional models were done with different power magnitudes, changing the nature of that power (i.e., injection of high-energy electrons or X-rays), or changing the injection profile. The choice of a constant power is a convenience. It overestimates the power in the case of a decreasing CSM density (see, e.g., \citealt{chevalier_fransson_94}) but the CSM density profile is not known a priori. Ultimately, this model set is a solid base to compare to observations and adjust the adopted properties of the injection power as suggested by the data.

The model nomenclature is such that logP3p35/mcds0p2/pwr1e40 corresponds to the progenitor logP3p35 (i.e., a primary star in a system with an original binary period of 10$^{3.35}$ or 2238\,d) from \citet{dessart_bin_24}, augmented with a CDS mass of 0.2\,\msun\ and with an injected power (in the form of high-energy electrons) of 10$^{40}$\,\ergs. Were this power injected in the form of X-rays, the name would be logP3p35/mcds0p2/xray1e40. When the power was injected away from the CDS centroid, the model name is appended with ``vloc$X$'' where $X$ is the new adopted velocity centroid (in \kms). All simulations started at 500\,d but the end time was often dictated by numerical difficulties (e.g., models with a lighter CDS were harder to converge at late times), and thus covers from $\sim$\,1000 to $\sim$\,2000\,d. Table~\ref{tab_set} summarizes the set of simulations we carried out in this work.

As stated at the start of this section, the radiative transfer modeling was done using the same approach as in \citet{dessart_23ixf_neb_26}. However, our model atom, which is identical for all simulations, includes higher ionization stages (specifically with stages \three\ and \four\ for CNO and intermediate mass elements) as well as a very large number of levels for iron-group elements. More ionization stages ensure that the gas can cool over a wide range of densities and temperature, which is essential at late times when the densities are low and when the gas is potentially energized by X-rays. A large model atom for iron-group element ensures that we are not underestimating the gas opacity, again for a wide range of densities and temperatures. Details on the model atom are given in Appendix~\ref{sect_atom}.

Finally, when comparing our models to observations, we sometimes invoke dust. The treatment of dust was adopted from \citet{dessart_dust_25} using a C-rich and Si-rich dust mixture as in the modeling of SN\,2023ixf \citep{dessart_23ixf_neb_26}. Dust properties (i.e., spatial distribution and amount) were given in the corresponding sections where dust is introduced, with the same nomenclature as in \citet{dessart_23ixf_neb_26}. For example, in the section describing a specific SN and a model, the dust characteristics labelled as ``ej/2em4/0p035/cds/1em4/0p035'' would refer to dust in the inner, metal-rich ejecta (typically below 2500\,\kms) with a temperature of 350\,K and a mass of $2 \times 10^{-4}$\,\msun, together with dust in the CDS with a temperature of 350\,K and a mass of $10^{-4}$\,\msun. At present, the dust is made of the same grain composition and size in a given calculation (only the spatial distribution and temperature can be modified).


\section{Dependence on time}
\label{sect_res_time}

\begin{figure}
\includegraphics[width=0.49\textwidth]{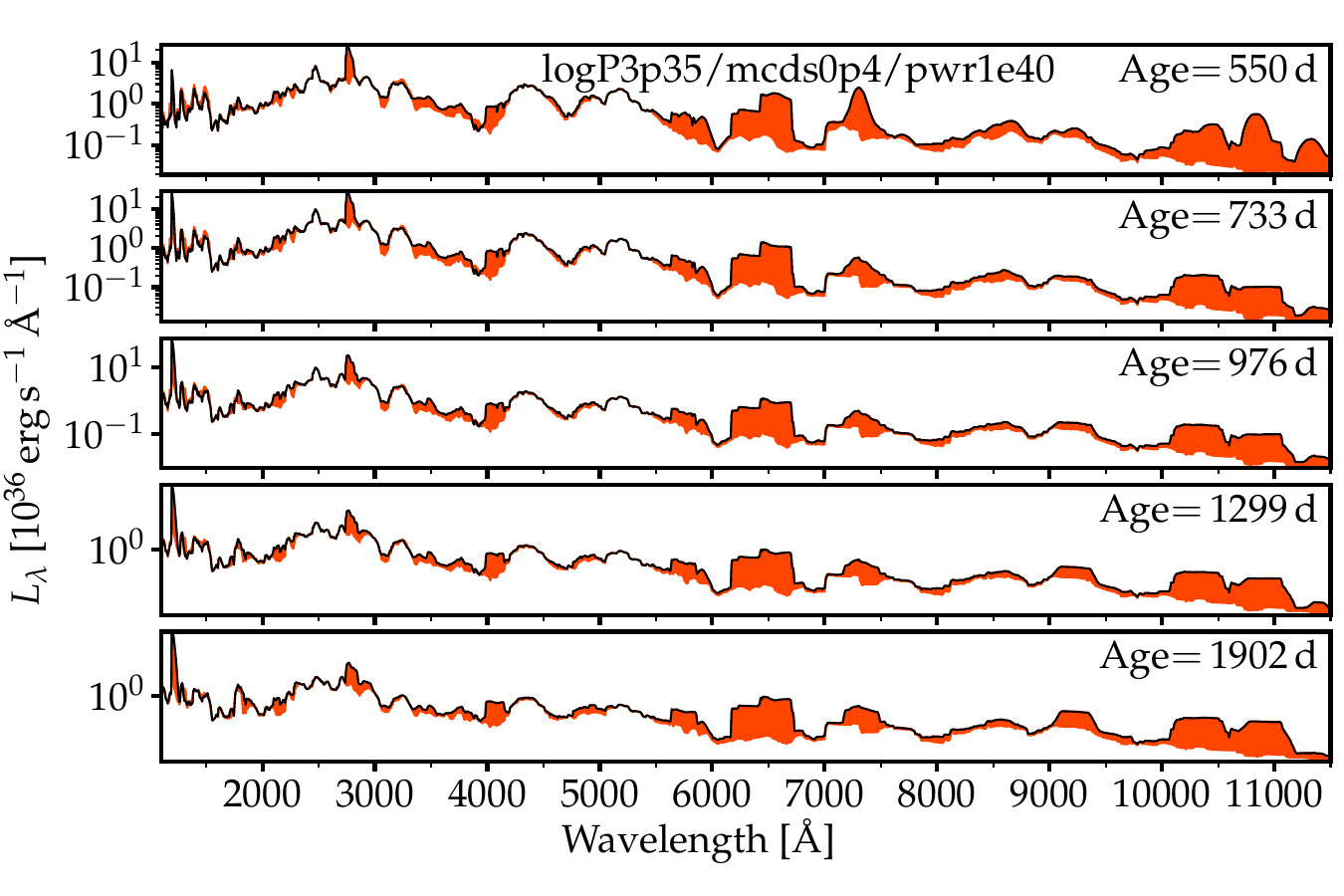}
\includegraphics[width=0.49\textwidth]{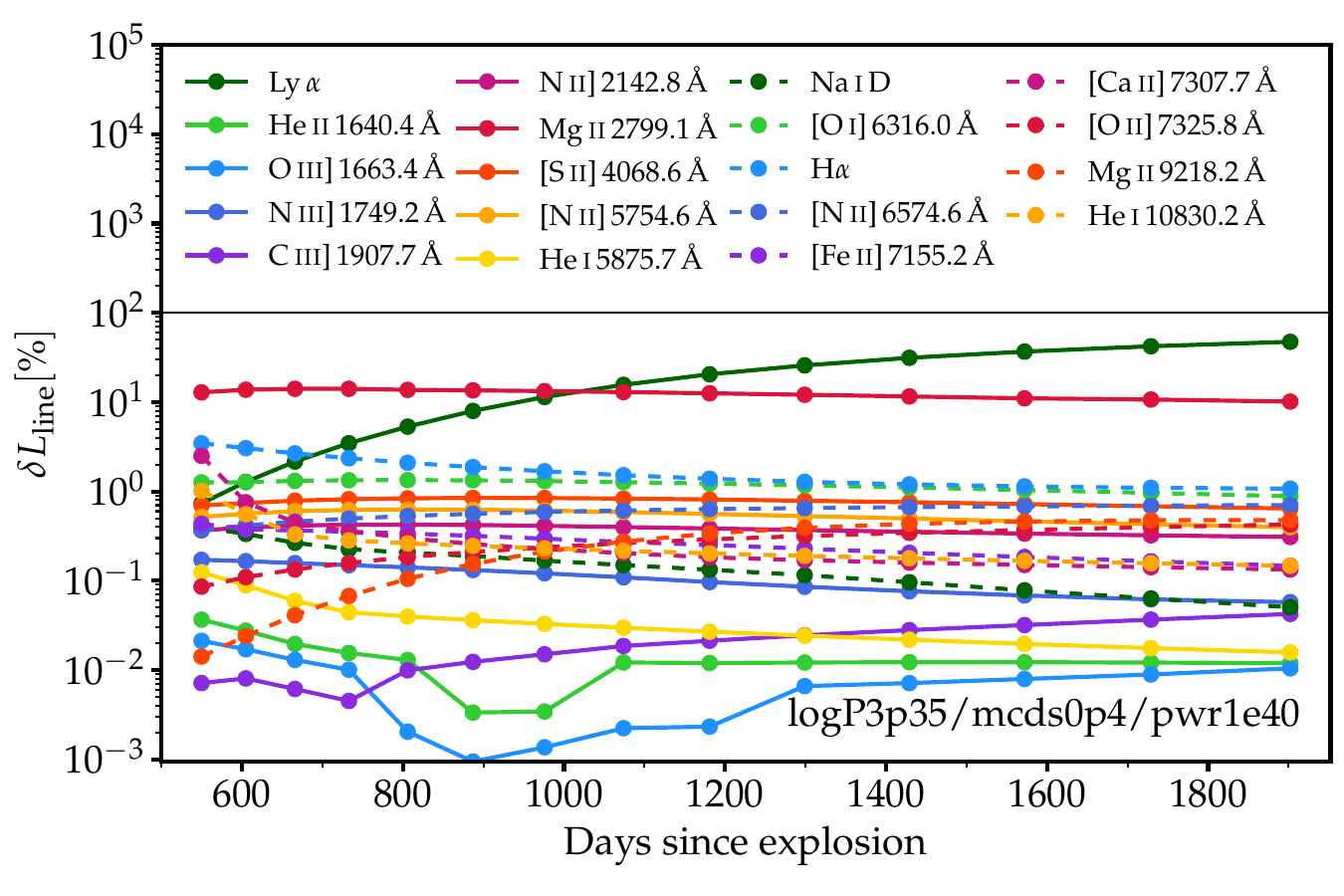}
\caption{Top: Spectral evolution from the UV to the optical of model logP3p35/mcds0p4/pwr1e40 from 550 until 1902\,d after explosion. The logarithm of the flux is shown to better reveal weak lines (the ordinate range covers several orders of magnitude). For each epoch and to emphasize the strong impact of iron on the escaping radiation, we shade in red the flux not associated with iron emission or absorption. Bottom: Corresponding evolution of specific line fluxes, given here as a percentage of the bolometric flux. Only a subset of models is shown in the upper panel. See Section~\ref{sect_res_time} for discussion.
\label{fig_logP3p35_evol}
}
\end{figure}

Figure~\ref{fig_logP3p35_evol} shows the evolution of the spectra as well as individual line fluxes for model logP3p35/mcds0p4/pwr1e40 from 550 until 1902\,d after explosion. Throughout most of the UV and optical, iron is the main element that sculpts the spectral appearance with only localized emission of stronger individual, generally forbidden lines. These lines are typically broad and boxy in the optical and infrared with a full Doppler width corresponding to twice the CDS velocity (\vcds\ is 6325\,\kms\ in this model). In the UV, all lines suffer strong optical depth effects and appear with a flux deficit that increases from the blue edge to the red edge. In some cases, this is caused by the intrinsically large optical depth of the line (e.g., \lya), but it also stems from the optical depth arising from other lines (e.g., forests of lines from Fe\two\ and Fe\three\ -- see also Fig.~21 of \citealt{dessart_23ixf_neb_26}). The Lyman continuum is highly optically thick across the CDS but the Balmer continuum is optically thin at 550\,d. Hence, in Fig.~\ref{fig_logP3p35_evol}, the noticeable narrowing of emission lines towards smaller wavelengths arises not only because Doppler broadening scales linearly with wavelength but also because of the increased profile skewness.

In the bottom panel of Fig.~\ref{fig_logP3p35_evol}, we show the fractional power radiated in individual emission lines. For the strong \lya\ line, this is measured from the total flux spectrum, but for weaker lines, this is measured from the model spectrum computed by accounting only for the associated species. This is necessary because of the severe optical depth effects, which prevent a simple calculation of those fluxes. Consequently, the line fluxes are indicative (this problem applies as well to real data from which the evaluation of line fluxes are very uncertain, in particular for emission lines in the UV where Fe absorption and emission impacts all lines). In this model at 550\,d, the strongest line in the spectrum is \mgiiuv, followed by \ha\ and \caiidoub\ and a myriad of other lines than carry one percent or less of the bolometric luminosity -- \lya\ carries only 0.73\,\%. As time passes, \mgiiuv\ remains roughly constant and the spectral energy distribution pivots around it as it shifts from optical to UV dominated. At 1902\,d, \lya\ has become the strongest line (i.e., 47.3\,\%) followed by \mgiiuv\ (10.1\,\%), \ha\ (1.1\,\%), \oidoub\ (0.9\,\%) an a myriad of other lines at 0.5\,\% or less. The strongest coolants in the CDS that carry away the injected power are therefore \lya\ and \mgiiuv. However, 74\,\% (34\,\%) of the flux is carried away by iron lines at 550\,d (1902\,d) so these iron lines are amongst the top coolants of the CDS (in this model the decay power represents only about a third of the shock power at 500\,d, about a tenth at 600\,d, and about a hundredth at 800\,d -- see also \citealt{dessart_late_23}).

Figure~\ref{fig_logP3p35_evol} includes numerous weak lines. There is generally a strengthening of forbidden lines by a factor of a few from 550 to 1902\,d but these remain subdominant relative to \lya, \mgiiuv\ and the forest of Fe lines. There is also a shift to higher ionization, illustrated for example by the strengthening of the contribution of \oiidoub\ in the 7300\,\AA\ region relative to \caiidoub. \niidoub\ and \niiauroral\ are both present at all times, the former contributing a sizable fraction of the emission around 6550\,\AA\ and the latter overlapping with \naid.

In the red part of the optical, the model shows a complex evolution from a mix of decay-powered (relatively narrow lines emitted from the inner ejecta) and shock-powered (relatively broad lines emitted from the CDS) emission lines to lines arising exclusively from the CDS. This includes \caiidoub\ which becomes increasingly broad and contaminated by \oiidoub\ at late times. At 550\,d and at longer wavelengths, the strong emission from \heinir\ overlaps with several lines of S\one\ and S\two\ as well as \pag, but this region becomes dominated at later times by a mix of broad boxy emission from Mg\two\,9231.2\,\AA, \siinir, \siiinir, \heinir, and Mg\two\,10914.2\,\AA.

\section{Dependence on CDS mass}
\label{sect_res_mcds}


%

\begin{figure}
\includegraphics[width=0.49\textwidth]{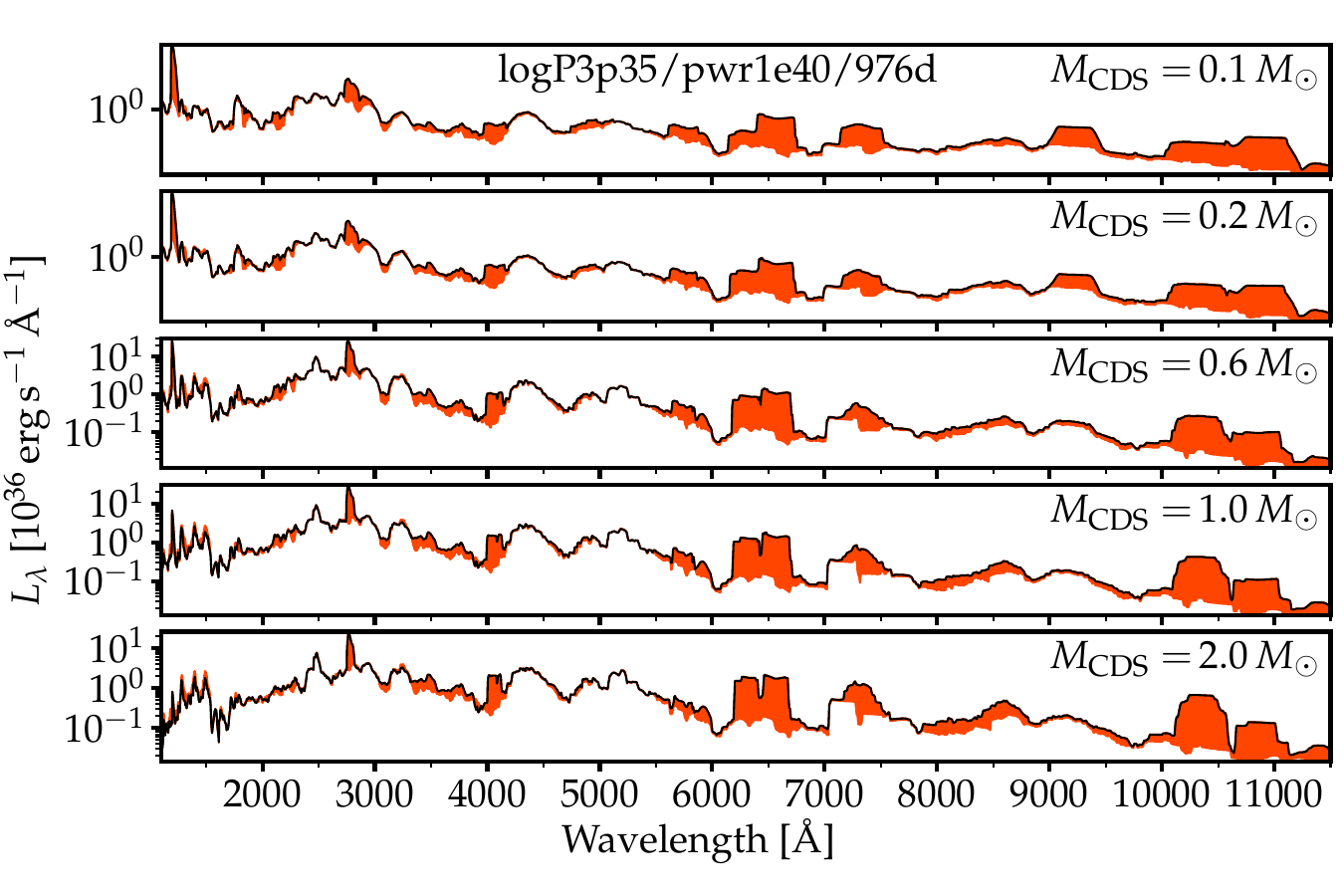}
\includegraphics[width=0.49\textwidth]{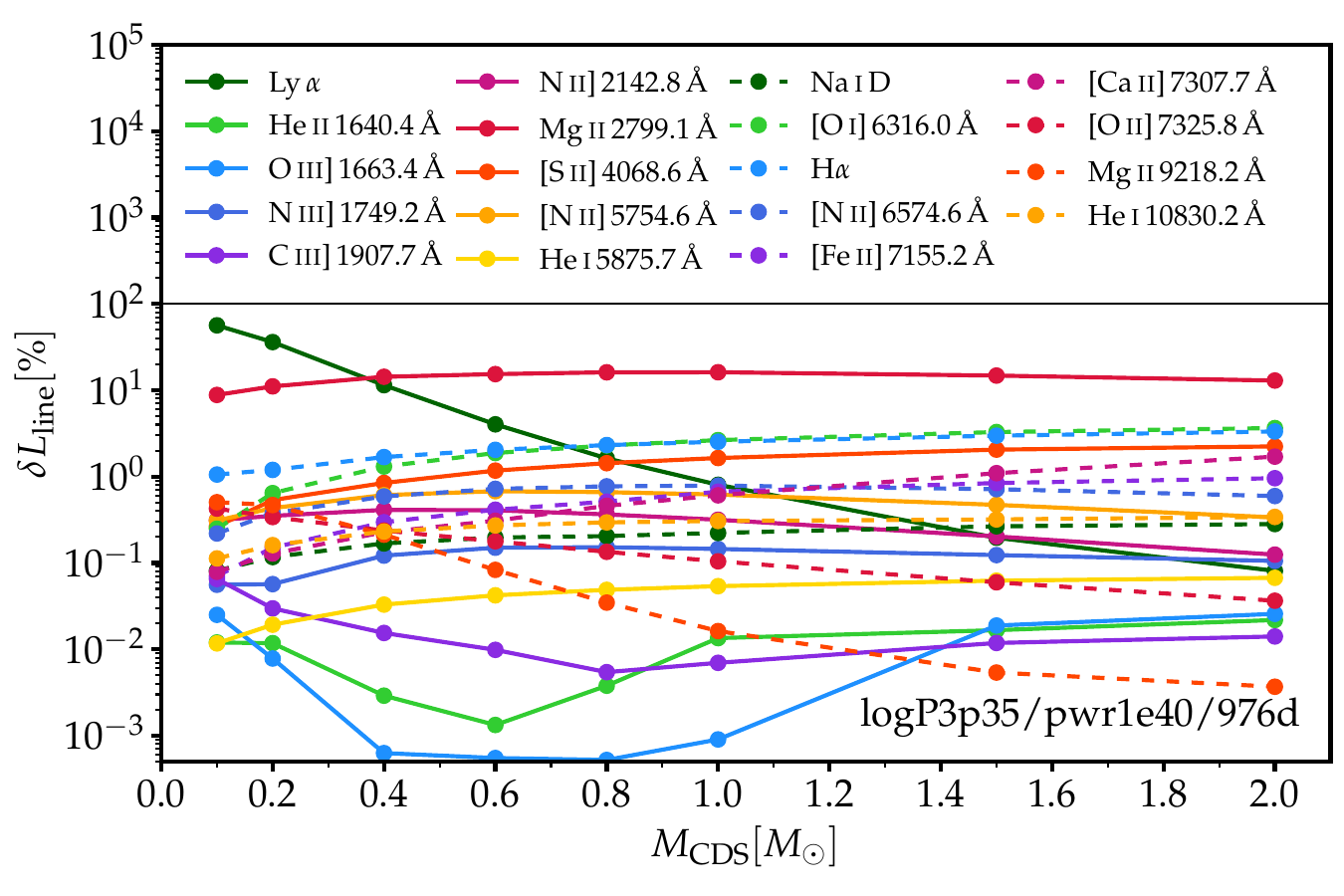}
\caption{Same as Fig.~\ref{fig_logP3p35_evol} but showing the results for model logP3p35 at 976\,d with a shock power of 10$^{40}$\,\ergs\ and augmented by a CDS of increasing mass from 0.1 to 2.0\,\msun.
\label{fig_logP3p35_mcds}
}
\end{figure}

Figure~\ref{fig_logP3p35_mcds} shows the evolution of the spectra as well as individual line fluxes when the CDS mass is increased from 0.1 to 2.0\,\msun\ -- we use model logP3p35 at 976\,d with the same power injection of 10$^{40}$\,\ergs\ for this comparison. This change in CDS mass also comes with a change in CDS velocity, since the more massive it is the slower it is (see Table~\ref{tab_set}). Hence, for higher CDS mass, the conditions in the CDS are denser, cooler, more optically thick, and the associated emission is less Doppler broadened. This is most strongly seen in the plunging of the \lya\ emission from the strongest line for \mcds\ of 0.1\,\msun\ to a negligible coolant for \mcds\ of 2.0\,\msun\ (in which case 71\,\% of the total cooling occurs through Fe emission). This drop is caused by the large CDS optical depth in \lya\ itself but also from the influence of overlapping Fe lines. \mgiiuv\ remains essentially constant across these values of \mcds.

The lower CDS ionization at higher \mcds\ leads to stronger lines from neutral species (e.g., \oidoub\ versus \oiidoub) and a weakening of transitions arising from higher excited states (e.g., Mg\,{\sc ii}\,9218.2\,\AA). In the models with the larger CDS velocity (i.e., lower \mcds), \ha\ and \oidoub\ overlap. This leads to an excess emission on the blue side of \ha\ although this excess is commensurate to the strength of \oidoub\ (specifically the red component of the doublet) -- this overlap is also clearly visible in the breakdown of species contributions in the model spectra shown in the appendix, for example those used for the comparison to SN\,2013ej. Overall, there is a relatively weak sensitivity of most of the lines to \mcds\ (the most sensitive line being \lya), which suggests that our results should not be overly compromised by the simplicity of our model setup.

\section{Dependence on progenitor H-rich envelope mass}
\label{sect_res_logP}


%

\begin{figure}
\includegraphics[width=0.49\textwidth]{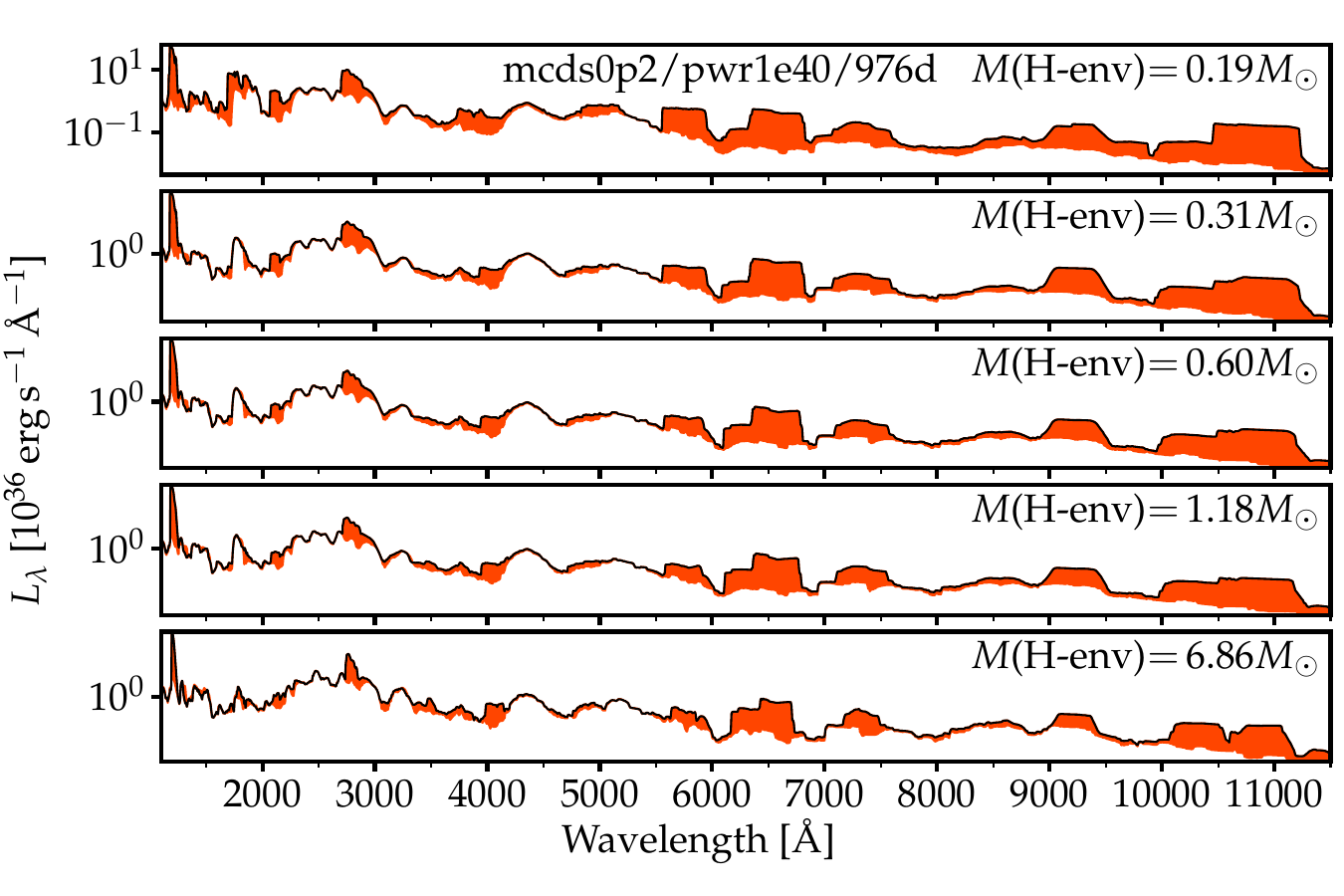}
\includegraphics[width=0.49\textwidth]{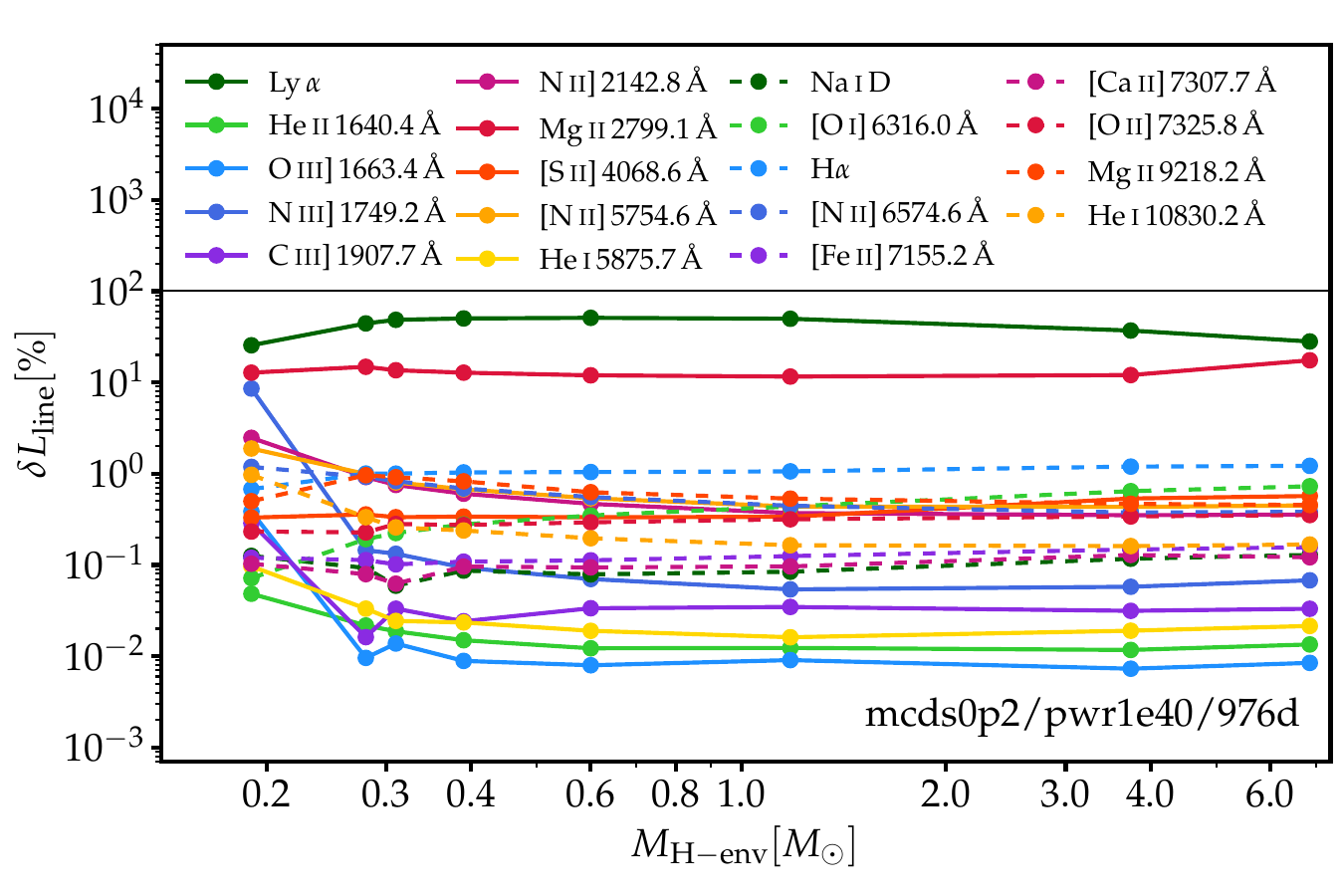}
\caption{Same as Fig.~\ref{fig_logP3p35_evol} but showing the results at 976\,d for a shock power of 10$^{40}$\,\ergs, a CDS mass of 0.2\,\msun, and for a set of progenitors endowed with an H-rich envelope mass between 0.1 and 6.86\,\msun\ at the time of explosion (i.e., models logP2p85 to logP3p45).
\label{fig_mcds0p2_logP}
}
\end{figure}

Figure~\ref{fig_mcds0p2_logP} shows the results for the set of progenitors endowed with an H-rich envelope between 0.1 and 6.86\,\msun\ at the time of explosion (i.e., models logP2p85 to logP3p45). The time is 976\,d after explosion, the shock power is 10$^{40}$\,\ergs, and the CDS mass is 0.2\,\msun\ for the whole set. As in the case of a varying \mcds, the lower is \mhenv\ the higher is the expansion rate of the ejecta and by extension that of the CDS (see Table~\ref{tab_set}). Hence, for the same \mcds, the CDS is of a lower density, which leads in our simulations to more ionized conditions in the CDS (for a fixed injected power). Furthermore, the gas is more He and N rich for lower \mhenv\ (see also Table~\ref{tab_set} and Fig.~\ref{fig_other_prog}) -- it covers a range of a factor of three at the progenitor surface in the set logP2p85 to logP3p45. Altogether, these variations affect the coolants, although mostly for the lowest values of \mhenv\ (i.e., less than 0.3\,\msun).

We find that \lya\ and \mgiiuv\ are the strongest coolants at all times (in competition with the bulk cooling effect of Fe lines) but their importance is modulated at low \mhenv\ because of an increased contribution from N lines (e.g., N\,{\sc iii}]\,1749.2\,\AA\ or \niidoub). The higher ionization at lower \mhenv\ also boosts the strength of \oiidoub\ and also leads to the clear presence of \oiiidoub\ and \neiiiopt\ (which overlap with Fe lines as well as \stwoopt) in model logP2p85. In the red part of the spectrum, there is a broad blend of \siinir, \heinir, and Mg\,{\sc ii}\,10914.2\,\AA. In all models, \ha\ and \oidoub\ remain strong in the optical.

\begin{table}
  \caption{Observational characteristics and data sources for the SN set used for model comparisons.
\label{tab_obs}
}
\begin{center}
\begin{tabular}{|l@{\hspace{1mm}}|c@{\hspace{1mm}}|c@{\hspace{1mm}}|c@{\hspace{1mm}}|c@{\hspace{1mm}}|c@{\hspace{1mm}}|}
\hline
SN name    &  $t_{\rm expl}$ & $cz$  & $D$   &     $E(B-V)$  & Ref. \\
              &   [MJD]        & [\kms]     & [Mpc] &       [mag]   &   \\
\hline
SN\,1993J    &  49074.0   &  33.9   & 3.63  & 0.2    &   a, b        \\    
SN\,1998S    &  50875.2   & 905.4   & 17.0  & 0.22   &   b, c, d     \\    
SN\,2013ej   &  56497.5   & 659.5   & 9.0   & 0.06   &   e, f        \\    
SN\,2023ixf  &  60082.8   & 241.0   & 6.85  & 0.041  &   g, h, i     \\    
\hline
\end{tabular}
\end{center}
Notes:
a: \citet{matheson_93j_00a}; b: \citet{fransson_cno_93J_98S_05};
c: \citet{fassia_98S_00}; d: \citet{leonard_98S_00};
e: \citet{yuan_13ej_16}; f: \citet{mauerhan_13ej_17};
g: \citet{jacobson_galan_23ixf_23};  h: \citet{bostroem_uv_26}; i: \citet{dessart_23ixf_neb_26}.
\end{table}


\section{Comparison to observations}
\label{sect_obs}

In this section, we use the grid of models described above to compare to UV (if available) and optical spectra of a sample of Type II SNe at late times of 1.5--3 years after explosion. We consider here in a chronological manner the SNe 1993J (Section~\ref{sect_res_93J}), 1998S (Section~\ref{sect_res_98S}), 2013ej (Section~\ref{sect_res_13ej}), and 2023ixf  (Section~\ref{sect_res_23ixf}). The SN characteristics used for each are given in Table~\ref{tab_obs}. In all comparisons, we corrected the observations for redshift and reddening and scaled the model to the SN distance. We generally plotted the flux $F_\nu$ rather than $F_\lambda$ because it yields a flatter flux distribution across the UV and optical and thus improves the visibility. A log scale was, however, used for the flux because of the large contrast in line strengths.



\begin{figure*}[h]
    \includegraphics[width=\textwidth]{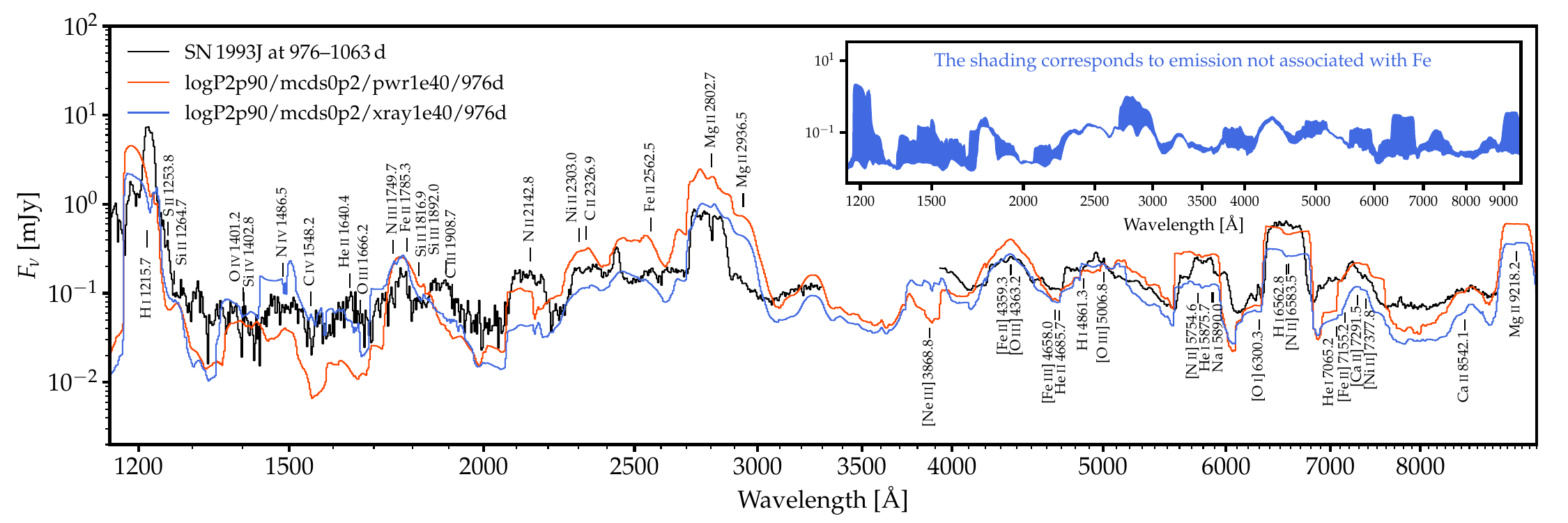}
\caption{Spectral comparison between the observations of SN\,1993J (at 1063\,d in the UV and at 976\,d in the optical; black) compared to the dust-free model logP2p90 with \mcds$=$\,0.2\,\msun\ and a power injection of $10^{40}$\,\ergs\ at 976\,d (as high-energy electrons [red] or as X-rays [blue]). The observations were corrected for reddening and redshift. The models were scaled to the distance of SN\,1993J. Labels indicate some of the strongest lines. The inset illustrates the emission that is not associated with Fe in the X-ray model. A complete description of the flux contribution from individual species is presented in Fig.~\ref{fig_93J_species} for the X-ray model.
\label{fig_93J}
}
\end{figure*}

\subsection{Comparison to SN\,1993J at 976--1063\,d}
\label{sect_res_93J}

  The Type IIb SN\,1993J has been extensively analyzed and modeled over the years. The light curve supports a heavily stripped progenitor with an extended $\sim$\,0.1\,\msun\ H-rich envelope at the time of explosion \citep{filippenko_93J_93,blinnikov_94_93j,nomoto_93j_93,dessart_bin_24}, compatible with a preSN evolution in an interacting binary \citep{podsiadlowski_93j_93}. This was independently confirmed by the identification of both the progenitor star as a K-type supergiant \citep{aldering_93j_94,vandyk_93j_02} and the existence of a companion star \citep{maund_93j_04,fox_93J_14}. At late times, interaction of the ejecta with CSM was inferred, for example, from the flattening of the SN optical brightness after about a year \citep{zhang_93j_late_04} but also from the sustained emission in the optical producing broad boxy line profiles \citep{matheson_93j_00a,houck_fransson_93J_96}, or from X-ray emission \citep{uno_93j_02,nymark_93J_09}. UV observations of SN\,1993J at late times revealed the presence of strong \lya, \mgiiuv, as well as a myriad of weaker emission lines from CNO suggesting N enrichment \citep{fransson_cno_93J_98S_05}.

In this section, we compare some of our models to the observations of SN\,1993J. We used an epoch around three years after explosion when both UV and optical data were collected. From our model set, we took a heavily stripped progenitor and selected a model with a CDS mass (and thus velocity) suitable to match the emission line widths observed in SN\,1993J. Figure~\ref{fig_93J} compares the UV data of \citet{fransson_cno_93J_98S_05} at 1063\,d\footnote{The UV data from HST were rereduced by K.A. Bostroem so that the \lya\ region could be extracted despite geocoronal contamination.} and the optical data of \citet{matheson_93j_00a} at 976\,d with a couple of models based on the progenitor logP2p90 with a CDS mass of 0.2\,\msun, an age of 976\,d and an injected power of 10$^{40}$\,\ergs. To test the sensitivity of the power injection and go beyond the set of models presented in earlier sections, we treated this power in the form of high-energy electrons (model pwr1e40) or in the form of X-rays (model xray1e40). In both cases, the power was injected with the same spatial distribution given by a gaussian with a centroid at 10\,492\,\kms\ (coinciding with the center of the CDS where the density is maximum) and a standard deviation of 35\,\kms. Because there is little evidence for dust in these observations, dust was ignored in the two models shown.

Figure~\ref{fig_93J} suggests a broad overall agreement of both models, driven by the fact that the bulk of the power is absorbed within the densest parts of the CDS which in both cases remains cool and partially ionized (i.e., $\sim$\,10\,000\,K; H, He, O neutral; Ca and Fe once ionized). However, in the X-ray case, the small fraction of X-rays that escape from the CDS and are absorbed above and below it cause a surge in temperature and ionization in those regions (i.e., 20--50\,000\,K; H fully ionized; He once and twice ionized; O three to four times ionized; Ca four times ionized; Fe five times ionized). In the model with a prescribed profile for the deposition of high-energy electrons, only the resulting radiation may photoionize the surrounding medium. This does not occur in the present case and the two models deviate consequently in a number of ways. In particular, the model with X-rays exhibits a number of high-ionization lines lacking in the pwr1e40 model, such as the O\four\ multiplet around 1400\,\AA, N\four\,1486.5\,\AA, C\four\,1550\,\AA, \neiiiopt\ or \oiiidoub. The flux in \lya\ and \mgiiuv\ are the strongest emission lines in SN\,1993J as well as in both models. A specificity of SN\,1993J are the strong N lines, in particular \niiauroral\ and \niidoub\ (which is nearly as strong as \ha\ in both models), which arises both from the relatively high ionization in the CDS (nitrogen is once ionized) and the N enrichment in this progenitor (see also Section~\ref{sect_res_logP} and Fig.~\ref{fig_mcds0p2_logP}). Another specificity of SN\,1993J is the broad and boxy appearance of many lines, not only in the optical but also in the UV (e.g., N\two]\,2142.8\,\AA, \mgiiuv). The optical range provides a good alternative to assess N enrichment relative to the UV \citep{fransson_cno_93J_98S_05}, which is unobservable from the ground and strongly affected by metal line blanketing even at late times.


\begin{figure}
\includegraphics[width=0.49\textwidth]{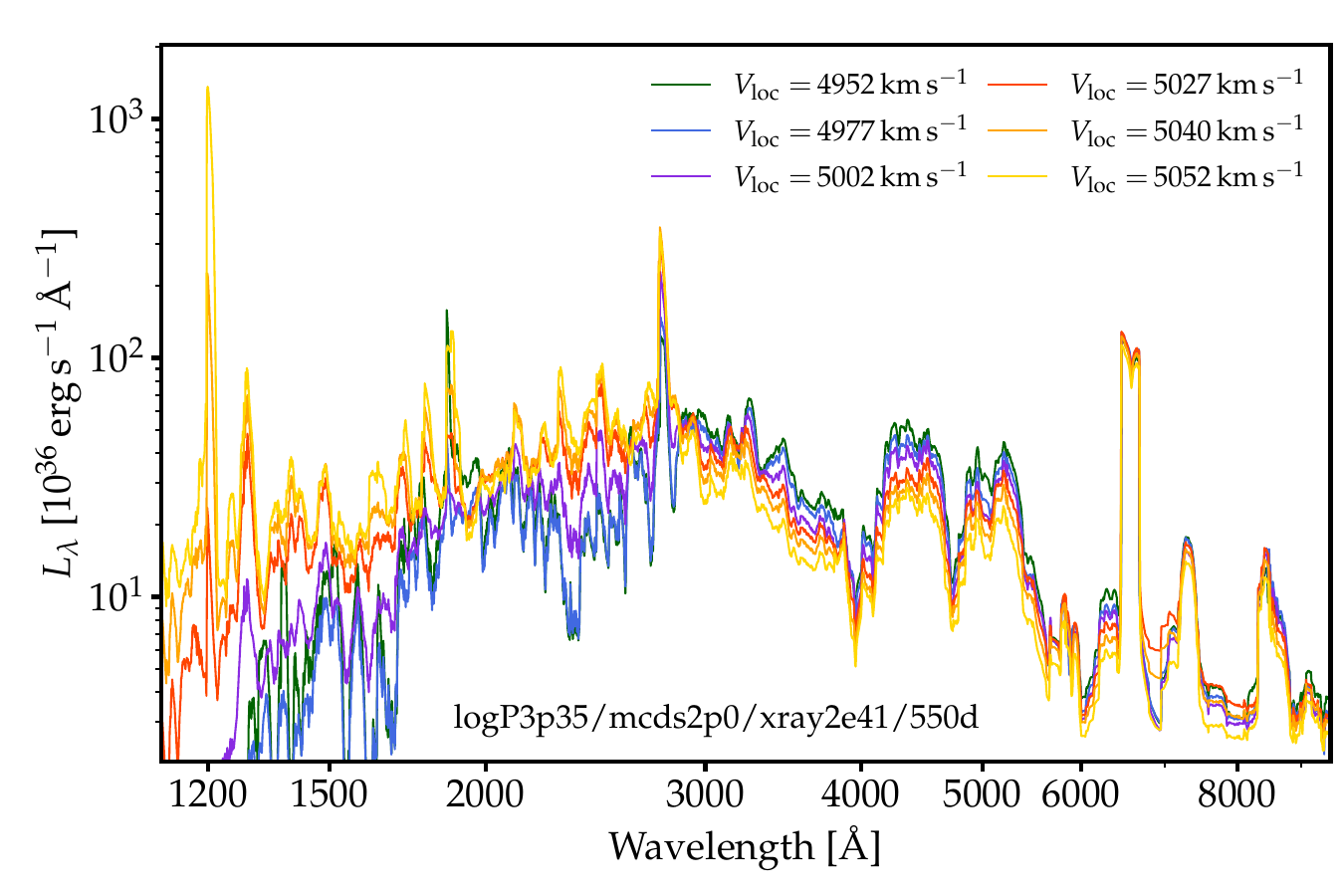}
\caption{Spectral variations caused by a shift of the power injection from a centroid at 5002\,\kms\ (nominal value corresponding the CDS peak density) to locations below and above the CDS in the dust-free model logP3p35/mcds2p0/xray2e41/550d shown in Fig.~\ref{fig_98S}.
\label{fig_98S_var_pwr}
}
\end{figure}


\begin{figure*}
\includegraphics[width=\textwidth]{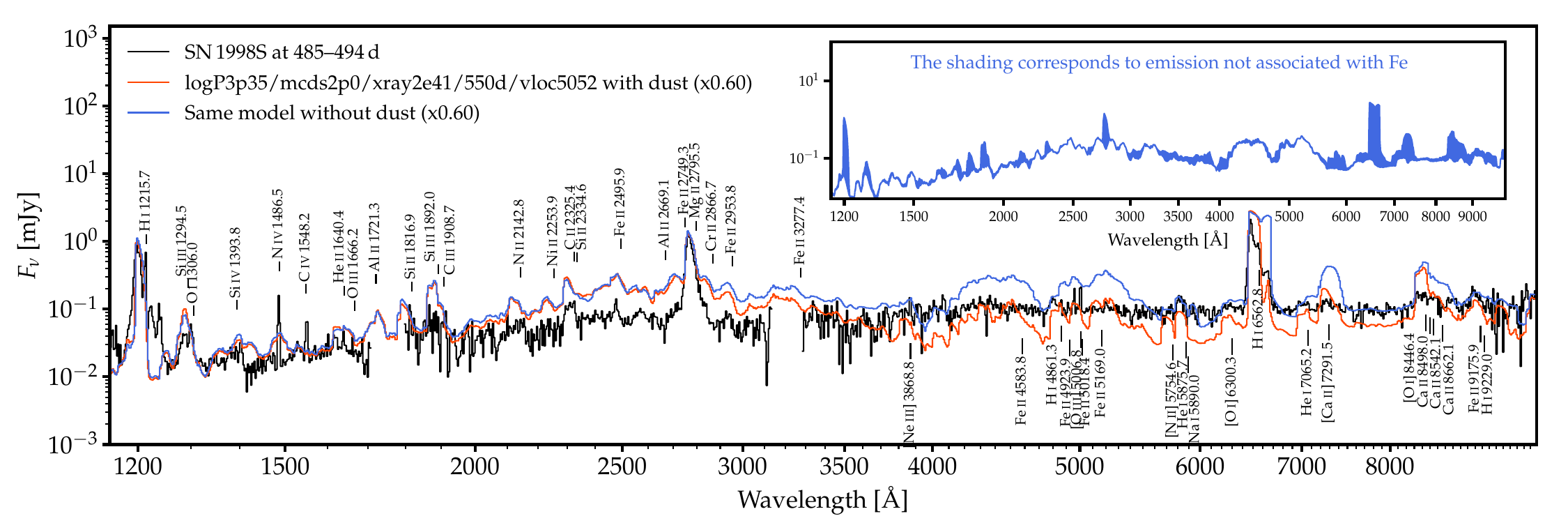}
    \caption{Spectral comparison between the observations of SN\,1998S at 485\,d (UV; \citealt{fransson_cno_93J_98S_05}) and 494\,d (optical; \citealt{leonard_98S_00}) and the model logP3p35/mcds2p0/xray2e41/550d/vloc5052 without dust (blue) and with dust (red: ej/2em4/0p035/cds/1em4/0p035). An additional scaling of 0.6 is applied to both models. A logarithmic scale is used for both wavelengths and fluxes. The inset shows the dust-free model together with its counterpart in which only iron was included, illustrating the regions [shaded] that are unrelated to iron. See Section~\ref{sect_res_98S} for discussion. A complete description of the flux contribution from individual species is presented in Fig.~\ref{fig_98S_species}.
\label{fig_98S}
}
\end{figure*}

\subsection{Comparison to SN\,1998S at 485--494\,d.}
\label{sect_res_98S}

We now turn to SN\,1998S and its long lived and multiple signatures of interaction with CSM. The presence of CSM in the immediate proximity of the exploding star led to the formation of emission lines with symmetric, electron-scattering broadened lines that survived for about a week \citep{leonard_98S_00,chugai_98S_01,fassia_98S_01,shivvers_98S_15,D16_2n,dessart_98S_25}. A percent-level polarization was also observed during this early interaction phase \citep{leonard_98S_00} and is compatible with an axisymmetric, extended, ionized, optically-thick CSM with a pole-to-equator density contrast of a few \citep{dessart_98S_25}. Interaction with CSM continued during the photospheric phase, as evidenced from the peculiar \ha\ line profile morphology with a strong and broad emission and a lack of P-Cygni absorption typical of standard Type II SNe \citep{D16_2n,HD19}. Nebular times saw the formation of molecules and dust, most likely in the CDS \citep{gerardy_nir_98S_00,pozzo_98S_04,dessart_dust_25}. UV observations also indicated persisting interaction at late times, with strong \lya\ and \mgiiuv\ emission as well as emission lines from CNO elements from which \citet{fransson_cno_93J_98S_05} inferred an elevated N enrichment.

In this section, we revisit these previous works, including our previous studies, and used our set of models to produce a rough description of SN\,1998S at about 490\,d when both UV and optical data are available. We used the optical data at 494\,d from \citet{leonard_98S_00} and the UV data at 485\,d from \citet{fransson_cno_93J_98S_05}. For the models, we selected one of the logP3p35 models with a CDS velocity that matched that of SN\,1998S, which is around 5000\,\kms. This meant using a CDS mass of 2.0\,\msun, which may be on the heavy side but as we found in Section~\ref{sect_res_mcds} the sensitivity to CDS mass is not large. We then adjusted the power injected to produce a model in line with the SN brightness, although it is mitigated by dust attenuation in this SN. Indeed, \citet{dessart_dust_25} found that the amount of dust needed to reproduce the optical line profile asymmetries in SN\,1998S also caused a $\sim$\,2\,mag extinction in the optical range. We employed a power injection of $2 \times 10^{41}$\,\ergs\ and in the form of X-rays.

When that power in the form of X-rays is injected within the CDS (i.e., with a gaussian profile centered at 5002\,\kms\ on the CDS and with a standard deviation of 35\,\kms), the \lya\ flux is essentially zero. Shifting the X-ray injection to regions below the CDS leads to even greater reduction, not just of \lya\ but of the entire UV flux, and in disagreement with the observations of SN\,1998S (Fig.~\ref{fig_98S_var_pwr}). This UV attenuation is caused by the large optical depth of the CDS, which is great in the UV and huge in \lya. Only models with an X-ray power injection exterior to the CDS produced a sizable \lya\ emission. In practice, to match the observations of SN\,1998S, we had to shift the centroid of the X-ray injection to 5052\,\kms, thus 50\,\kms\ faster than the CDS centroid. With this choice, the outer parts of the CDS produce the entire emission from the SN. This arises mostly from the fact that the column density is smaller from such regions, leading to reduced optical depth. But it also arises from the enhanced ionization in those regions, which shuts off blanketing effects from Fe\two\ or Fe\three\ (see also discussion in \citealt{dessart_23ixf_neb_26}). Indeed, the outer regions of the CDS are then dominated by ions like He$^{+}$, O$^{2+}$, or Fe$^{3+}$. With our assumption of homologous expansion, the model thus predicts broad emission lines from such ionized species -- these predicted lines are also blueshifted because of optical-depth effects. In the observations, there is clear emission from N\four\,1486.5\,\AA, C\four\,1550\,\AA\ or He\two\,1640\,\AA, but the lines are narrow and indicates their formation in the unshocked CSM. There is thus some physical basis in the model predictions in terms of power injection and ionization stratification, the main limitation of the code in this sector being the inadequate assumption of homology which leads to the prediction of broad emission when sometimes it is instead narrow.

Figure~\ref{fig_98S} compares the observations of SN\,1998S at 485\,d (UV; \citealt{fransson_cno_93J_98S_05}) and 494\,d (optical; \citealt{leonard_98S_00}) with the model exhibiting strong \lya, \mgiiuv, and \ha, namely model logP3p35/mcds2p0/xray2e41/550d/vloc5052. The figure includes both the model without dust (blue) and with dust (red: ej/2em4/0p035/cds/1em4/0p035). The agreement is good for \lya, \mgiiuv, and \ha, in particular with a good match to the peak flux of these three landmarks. There is however significant offsets for the weaker features although most of the observed emission lines are also present in the model. In the optical, He\one\ lines at 5875 and 7065\,\AA\ are predicted, together with \caiidoub\ (\oiidoub\ is very weak in the model) -- the skewness of the profiles caused by dust complicates the identification. The Ca\two\ NIR triplet is contaminated by a strong O\one\,8446.5\,\AA, which is unusual at such late epochs but arises from the high density in the CDS. What is lacking is the presence of extended emission from Fe\two\ in the optical, which is very strong in the model not just in the optical but also in the UV -- the entire spectrum is essentially an iron spectrum if one ignores the three strongest lines (\lya, \mgiiuv, and \ha).

The model with dust reproduces roughly the blue-red asymmetry of \ha\ with $2 \times 10^{-4}$\,\msun\ of dust in the inner ejecta and $10^{-4}$\,\msun\ of dust in the CDS -- this is a lower dust mass than obtained in \citet{dessart_dust_25} because we adopted here a mixture of C-rich and Si-rich dust (same choice as in \citealt{dessart_23ixf_neb_26}) whereas we used Silicate dust only in \citet{dessart_dust_25}. Invoking dust to explain the \ha\ profile asymmetry may be an oversimplification since a reversed profile skewness is observed in the interacting SN PTF11iqb \citep{smith_ptf11iqb_15} -- both dust and asymmetry (of either or both ejecta and CSM) may be involved. The introduction of dust has little impact on the UV in this matching model for SN\,1998S (Fig.~\ref{fig_98S}).



\begin{figure}
\includegraphics[width=0.49\textwidth]{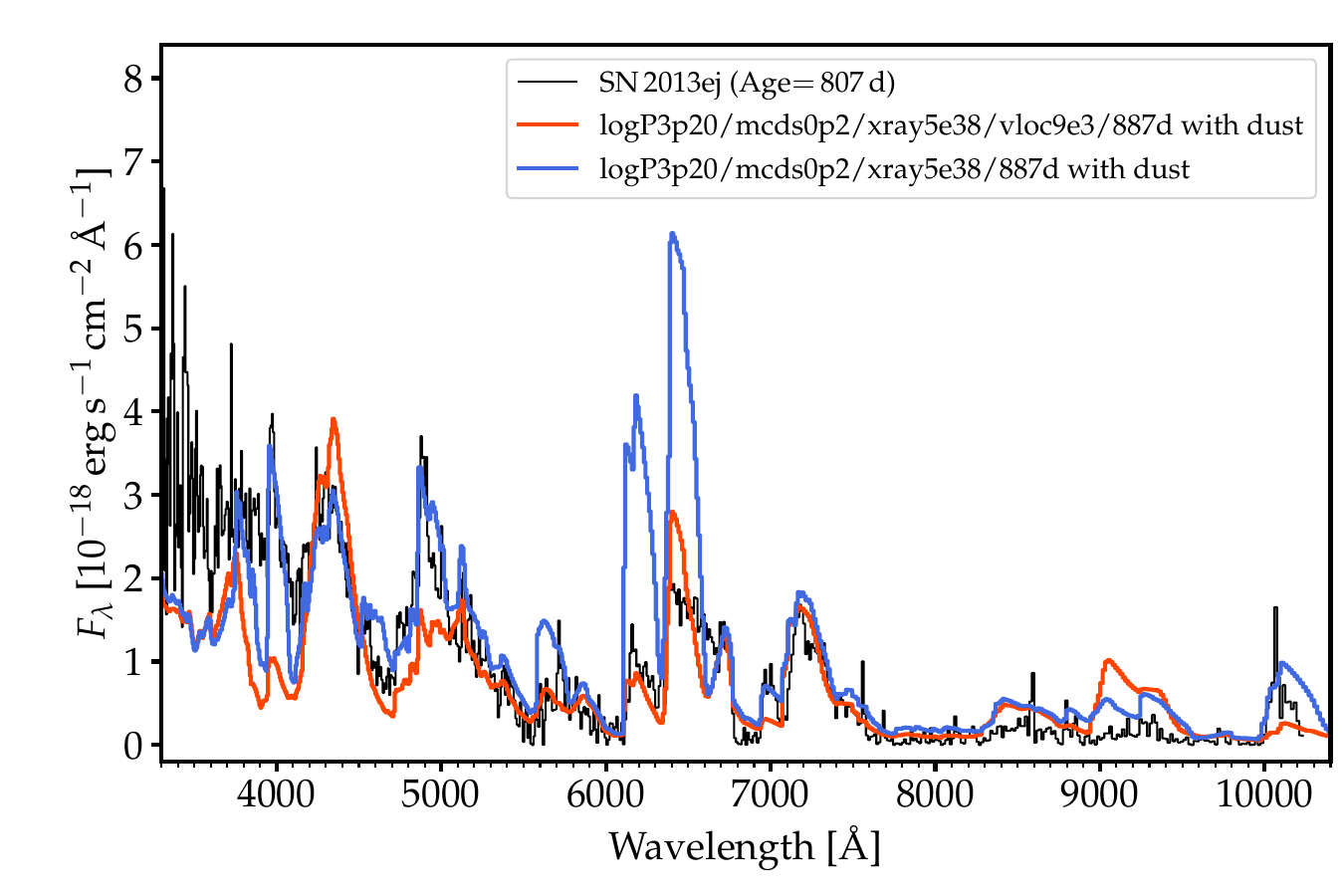}
\caption{Spectral comparison between the observations of SN\,2013ej at 807\,d (black; \citealt{mauerhan_13ej_17})  from and the model logP3p20 at 887\,d with \mcds$=$\,0.2\,\msun\ and an injection of X-rays with a total power of $5 \times 10^{38}$\,\ergs\  at 9231\,\kms\ (blue curve; this corresponds to the CDS velocity) or at 9000\,\kms\ (red curve; model whose name contains vloc9e3). The observations have been corrected for reddening and redshift, whereas the model has been scaled to the distance of SN\,2013ej. In both cases, a mixture of 350\,K Si-rich and C-rich 1.0\mic\ dust grains is assumed with $5 \times 10^{-3}$\,\msun\ uniformly distributed below 2500\,\kms\ and $4 \times 10^{-3}$\,\msun\ within the dense shell at 9231\,\kms. Models and observations have been rebinned to a fixed resolution of 500\,\kms. A complete description of the flux contribution from individual species for both models is presented in Fig.~\ref{fig_13ej_species}.
\label{fig_13ej}
}
\end{figure}

\subsection{Comparison to SN\,2013ej at $\sim$\,807\,d.}
\label{sect_res_13ej}

The Type II SN\,2013ej is characterized by a long rise time, a short photospheric phase, and a steep decline during the photospheric phase \citep{valenti_13ej_14,bose_13ej_15}. \citet{utrobin_chugai_13ej_17} proposed a large ejecta mass of 23--25\,\msun\ with a standard \ekin\ but high-velocity \nifs. Interaction with CSM is, however, likely at the origin of a number of peculiarities observed in SN\,2013ej. The presence of CSM in the immediate environment of the progenitor improves the match to the light curve during the photospheric phase \citep{morozova_2l_2p_17}. Using a $1.2 \times 10^{51}$\,erg $\sim$\,11\,\msun\ ejecta with early CSM interaction, \citet{HD19} could match the multi-band light curves and spectra of SN\,2013ej during the photospheric phase, including the peculiar line profile shapes (e.g., \ha) -- the weak absorption relative to the emission is one signature of previous or ongoing interaction with CSM in SNe II (see also \citealt{gutierrez_ha_14}; \citealt{dessart_csm_22}; \citealt{csornyei_gutierrez_26}). Interaction with CSM was independently supported from X-ray detection \citep{margutti_xray_13ej_13}. Modeling of nebular-phase observations suggest a 12-15\,\msun\ progenitor \citep{yuan_13ej_16,dessart_sn2p_21}, also compatible with inferences based on pre-explosion photometry \citep{fraser_13ej_14}. Remarkably, SN\,2013ej continued to be detectable at very late times of a few years, indicative of ongoing interaction with CSM, together with infrared thermal emission associated with dust formation \citep{mauerhan_13ej_17}.

Figure~\ref{fig_13ej} compares the observations of SN\,2013ej at 807\,d (black; \citealt{mauerhan_13ej_17}) and the model logP3p20 at 887\,d with \mcds$=$\,0.2\,\msun\ and an injection of X-rays with a total power of $5 \times 10^{38}$\,\ergs\ either centered at 9231\,\kms\ (blue curve; this corresponds to the maximum density in the CDS) or below the CDS at 9000\,\kms\ (red curve; model whose name contains vloc9e3). In both cases, the power injection follows a gaussian profile with a standard deviation of 35\,\kms. A mixture of 350\,K Si-rich and C-rich 1.0\mic\ dust grains is used in both models with $5 \times 10^{-3}$\,\msun\ uniformly distributed below 2500\,\kms\ and $4 \times 10^{-3}$\,\msun\ within the dense shell at 9231\,\kms. Neither model matches all features but each matches well a subset of features.

The model with X-ray injection within the CDS overestimates the strength of \oidoub\ and \ha\ but yields a satisfactory match to most other regions, including those around \stwoopt, \neiiiopt, \oiiidoub, He\one\,7065\,\AA, or \caiidoub. The model with the deeper X-ray injection matches better  \oidoub\ and \ha\  but yields a worse match of the \neiiiopt\ and \oiiidoub\ region. These differences arise from the different impact of the dust located in the CDS relative to power injection within the CDS or below. If dust is external to the emission, it produces no blue-red asymmetry in the associated emission line profiles. So, while the intrinsic \oiiidoub\ emissivity is comparable in both models, the dusty model with X-ray injection within the CDS yields a better match of that feature because the blue-red asymmetry arises from the optically-thick dust in the CDS. The other model shows a flat emission around 5000\,\AA\ (a large part of this emission comes from Fe, as for the 4300\,\AA\ bump) because this emission comes from below the dust-rich CDS. This exercice shows how difficult it is to match the whole spectrum consistently and indicates that inferences of dust properties from individual profiles can be problematic since unblended lines are rare below 5500\,\AA.




\begin{figure*}
\includegraphics[width=\textwidth]{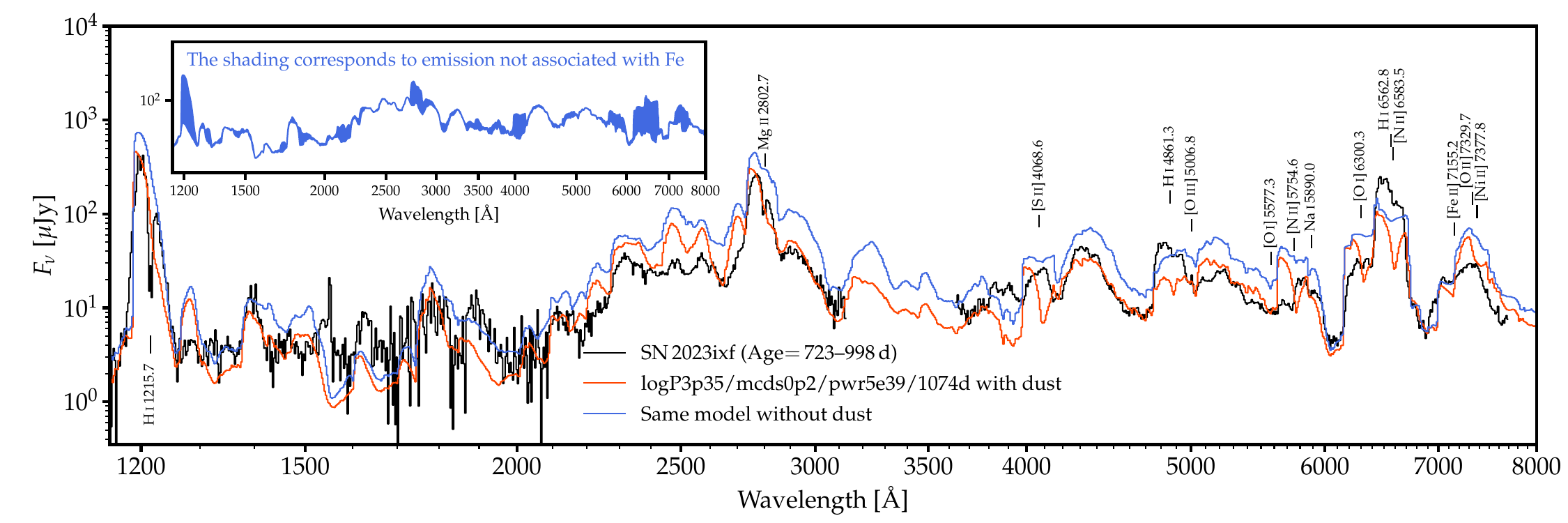}
\caption{Spectral comparison between the observations of SN\,2023ixf (at 723\,d in the UV and at 998\,d in the optical; black) compared to the model logP3p35 with \mcds$=$\,0.2\,\msun\ and an injected power of $5 \times 10^{39}$\,\ergs\ at 1074\,d with dust (red) and without dust (blue). The observations have been corrected for reddening and redshift, whereas the model has been scaled to the distance of SN\,2023ixf. For the dusty model, a mixture of 350\,K Si-rich and C-rich 0.1\mic\ dust grains is assumed with $5 \times 10^{-4}$\,\msun\ uniformly distributed below 2500\,\kms\ and $2 \times 10^{-4}$\,\msun\ within the dense shell at 6970\,\kms. Models and observations have been rebinned to a fixed resolution of 500\,\kms. Labels indicate some of the strongest lines. The inset illustrates the emission that is not associated with Fe in the dust-free model. A complete description of the flux contribution from individual species is presented in Fig.~\ref{fig_23ixf_species}.
\label{fig_23ixf}}
\end{figure*}

\subsection{Comparison to SN\,2023ixf at 723--998\,d}
\label{sect_res_23ixf}

Previous work on nebular phase observations of SN\,2023ixf clearly indicated that the SN was essentially entirely powered by interaction after about two years. This arose from the progressive strengthening of the UV relative to the optical \citep{wynn_sed_25,bostroem_uv_26,dessart_23ixf_neb_26}, but also from the dominance of broad-boxy emission lines in the optical at later times relative to the dominance of narrow emission lines from the decay-powered inner ejecta \citep{folatelli_23ixf_25,kumar_23ixf_25,zheng_23ixf_25,dessart_23ixf_neb_26}, in close correspondance to predictions \citep{dessart_late_23}.

In this section, we extend the previous work of \citet{dessart_23ixf_neb_26} and present simulations based on model logP3p35 with a CDS mass of 0.2\,\msun\ and velocity of 6966\,\kms\ (which is more suitable to match the widths of emission lines like \ha\ than the fixed value of 8000\,\kms\ used for all nebular epochs in \citealt{dessart_23ixf_neb_26}). Using this model as a reference, numerous attempts were carried out by varying the interaction power (its magnitude, location, as well as its nature as high-energy electrons or X-rays) and the dust properties. For the latter, the amount of dust in the inner ejecta is largely irrelevant for the UV and optical range since most of that radiation at $\gtrsim$\,2\,yr arises from the outermost ejecta regions (i.e., around the CDS). So, as far as the UV and optical are concerned, there can be as much inner-ejecta dust as one wishes -- we adopted here $5 \times 10^{-4}$\,\msun\ uniformly distributed below 2500\,\kms.

Figure~\ref{fig_23ixf} shows a comparison between the UV (at 723\,d) and optical (at 998\,d) observations of SN\,2023ixf and the model logP3p35/mcds0p2/pwr5e39 at 1074\,d with and without dust -- the slow evolution in the UV at those late times justifies merging data with such a large time offset (Jacobson-Gal\'an et al., in prep.). The dust in the CDS at 6970\,\kms\ corresponds to $2 \times 10^{-4}$\,\msun\ of a mixture of 350\,K Si-rich and C-rich 0.1\mic\ grains. As before \citep{dessart_dust_25,dessart_23ixf_neb_26}, we find that the dust has a moderate impact on the UV because it forms in the outer parts of the CDS and thus largely external to the dust-rich regions -- this impact is less obvious in the UV because UV lines exhibit a blue-red asymmetry even in the absence of dust. But it impacts the optical significantly (where most lines are optically thin in the absence of dust), with a blue-red asymmetry as well as a central dip in emission lines that arises from a greater attenuation from the limbs of the CDS (this feature is in part an artefact of the assumed spherical symmetry). The inset indicates that most of the spectrum is shaped by iron emission and absorption. Exceptions are \lya, N\,{\sc ii}]\,2142.8\,\AA, \mgiiuv,
\stwoopt, \neiiiopt\ (underestimated by the model, although the data show no evidence for \oiiidoub), \niiauroral, \naid, \oidoub, \ha, \feiiopt, \caiidoub, \oiidoub, and \nkiiopt. The match of the dusty model is somewhat superior to that presented in \citet{dessart_23ixf_neb_26} mostly because of the reduced CDS velocity (in that paper, the optical observations at 998\,d were analyzed on their own).


\section{Conclusions}
\label{sect_conc}

In this work, we revisited our previous late-time radiative-transfer studies of ejecta interaction with CSM \citep{dessart_csm_22,dessart_late_23} by considering a wide range of red-supergiant progenitors endowed with an H-rich envelope mass spanning from 0.19 to 6.86\,\msun\ as a result of binary mass exchange. In the absence of CSM, such explosions would produce SNe with properties typical of Type IIb or Type II with various decline rates from II-L to II-P \citep{dessart_bin_24}. The main motivation for this work was to produce models capable of reproducing the UV and optical data of emblematic, well-observed, nearby SNe past the age of one year, which is the representative epoch for nebular phase observations or models for Type II SNe, and extend this to several years. Such (published and public) data exist for our selected sample, which includes SNe 1993J, 1998S, 2013ej, and 2023ixf.

To alleviate the difficulty of modeling such ejecta at late times of 500 to 2000\,d, we made a number of simplifications by adopting ejecta with a smooth density structure, a mixed composition, and a fixed \nifs\ mass of 0.08\,\msun. These ejecta were augmented with a CDS covering a range of masses (0.1 to 2.0\,\msun) and velocities (5002 to 10643\,\kms) suitable for weak to strong interactions in this progenitor set. The shock power was treated in the form of high-energy electrons or X-rays injected with a gaussian profile in or around the CDS.  The emerging radiation from such interacting ejecta encodes the complexity inherent to core-collapse SN ejecta (i.e., progenitor mass and composition, explosive nucleosynthesis, chemical mixing), ejecta interaction with CSM (mass, velocity, as well as small and large-scale structure of the CDS), but also the influence of dust that may be both intrinsic (forming in the inner metal-rich ejecta as well as in the CDS) and external. Obtaining a good match to observations throughout the UV and optical is thus a considerable challenge.

Our late-time spectra are generally dominated by \lya, \mgiiuv, and \ha. As a fraction of the total flux, \mgiiuv\ changes little over time. Because of optical depth effects, \lya\ strengthens considerably with increasing time or with decreasing \mcds. With increasing time, we tend to see a strengthening of forbidden lines and an increase in ionization. Conditions in the CDS are typically warm and partially ionized, but any energy injection above or below the CDS produces emission from high-ionization lines, with for example \oiiidoub. Although Fe contributes only few strong isolated lines (e.g., \feiiopt), it contributes a large share of the total flux, in particular below 5500\,\AA\ where most of the flux resides at these late times. Overall, lines appear broad and boxy in the optical, but tend to show a strong red deficit in the UV, even in the absence of dust.

\lya\ is present in observations of a wide range of interacting transients \citep{bostroem_uv_26}, even at relatively early times (e.g., SN\,1998S; \citealt{fransson_cno_93J_98S_05}), whereas in models this line shows a strong sensitivity to the CDS mass and velocity. This sensitivity arises from the large background opacity brought in by metal lines and in particular from Fe but also from the huge optical depth in \lya\ itself.
Hence, we find that any \lya\ emission from below and within the CDS is strongly attenuated (even in the absence of dust), and that instead \lya\ emission arises from the outer parts of the CDS. In events in which dust is identified in the optical from a blue-red asymmetry in lines like \ha, \lya\ is still present and well matched by a dust-free model (e.g., SN\,1998S; Section~\ref{sect_res_98S}). This is thus independent evidence that the UV flux only forms and escapes from the outer parts of the CDS, even in the absence of dust, and not from regions underneath the CDS and influenced by the reverse shock. This finding is based on the models presented here, which cover a limited range in post-explosion epochs (at most 500 to 1900\,d)  and the few SN observations these models were confronted to. At later times of decades post-explosion, this situation will likely differ.

Our models with a higher He and N abundance in the H-rich material (models with lower \mhenv) exhibit stronger N lines relative to C and O lines, in particular with the most stripped progenitor (model logP2p85). This is seen in the UV with, for example, N\two]\,2142.8\,\AA\ and N\three\,1749.7\,\AA, but also in the optical with the enhanced flux in \niiauroral\ and \niidoub. The latter overlaps and sometimes rivals with \ha\ in our models. Such N lines are observed in SN\,1993J and also roughly matched with our model logP2p90 (progenitor with a 0.28\,\msun\ H-rich envelope) augmented with a CDS of 0.2\,\msun.

O\three\ lines arise in our simulations when some shock power is absorbed either above or below the CDS. Such lines do not form in the dense shell at those epochs because of its relatively high density and low ionization. If such lines form below the CDS (i.e., in the ejecta), any dust in the CDS yields a globally attenuated line emission and no profile skewness results. If it forms outside the CDS, dust absorption within the CDS skews the profile with a flux deficit redward of line center. The identification of these effects is challenged by the overlap of such lines with the background flux from Fe, which sometimes dominates in the given spectral region. Our dusty models yield a good match to those features in SN\,2013ej.

The two nearby SNe 2023ixf and 2024ggi offer an opportunity to study the long-term evolution of ejecta interaction with CSM across the electromagnetic spectrum. Arising from progenitors endowed with a larger mass at core collapse, they will offer an instructive contrast to other well observed events like SN\,1993J.


\begin{acknowledgements}
We acknowledge fruitful discussions with Norbert Langer and Andrea Ercolino. This work was granted access to the HPC resources of TGCC under the allocation 2025 -- A0190416871 made by GENCI, France. This research has made use of NASA's Astrophysics Data System Bibliographic Services.
\end{acknowledgements}


\bibliographystyle{aa}
\bibliography{./new_sn_library_luc}

\appendix

\onecolumn

\section{Model atom}
\label{sect_atom}

The species (atoms and ions) included in all calculations are
H\one\ (26,36),
He\one\ (40,51), He\two\ (13,30), C\one\ (14,26), C\two\ (14,26), C\three\ (62,112), C\four\ (59,64),
N\one\ (44,104), N\two\ (23,41), N\three\ (25,53), N\four\ (56,92),
O\one\ (21,51), O\two\ (54,123), O\three\ (44,86), O\four\ (76,152),
Ne\two\ (22,91), Ne\three\ (32,80), Ne\four\ (30,67),
Na\one\ (22,71),
Mg\one\ (39,122), Mg\two\ (31,80), Mg\three\ (31,99), Mg\four\ (18,100),
Al\two\ (26,44), Al\three\ (27,60),
Si\one\ (100,187), Si\two\ (31,59), Si\three\ (33,61), Si\four\ (37,48),
S\one\ (106,322), S\two\ (56,324), S\three\ (48,98), S\four\ (27,67),
Ar\one\ (56,110), Ar\two\ (134,415), Ar\three\ (32,346),
K\one\ (25,44),
Ca\two\ (21,77), Ca\three\ (16,40), Ca\four\ (18,69),
Sc\two\ (38,85), Sc\three\ (25,45),
Ti\two\ (37,152), Ti\three\ (33,206),
Cr\two\ (28,196), Cr\three\ (30,145), Cr\four\ (29,234)
Fe\one\ (413,1142), Fe\two\ (228,2698), Fe\three\ (96,1001), Fe\four\ (100,1000), Fe\five\ (139,1000),
Co\two\ (112,1005), Co\three\ (88,1075), Co\four\ (56,1000),
Ni\one\ (56,301), Ni\two\ (59,1000), Ni\three\ (47,1000), and Ni\four\ (54,1000).
The numbers in parenthesis correspond to the number of super-levels and full levels -- for details see \citet{HM98}.

\section{Additional illustrations}
\label{sect_more}

\begin{figure*}
\begin{center}
\includegraphics[width=0.7\textwidth]
{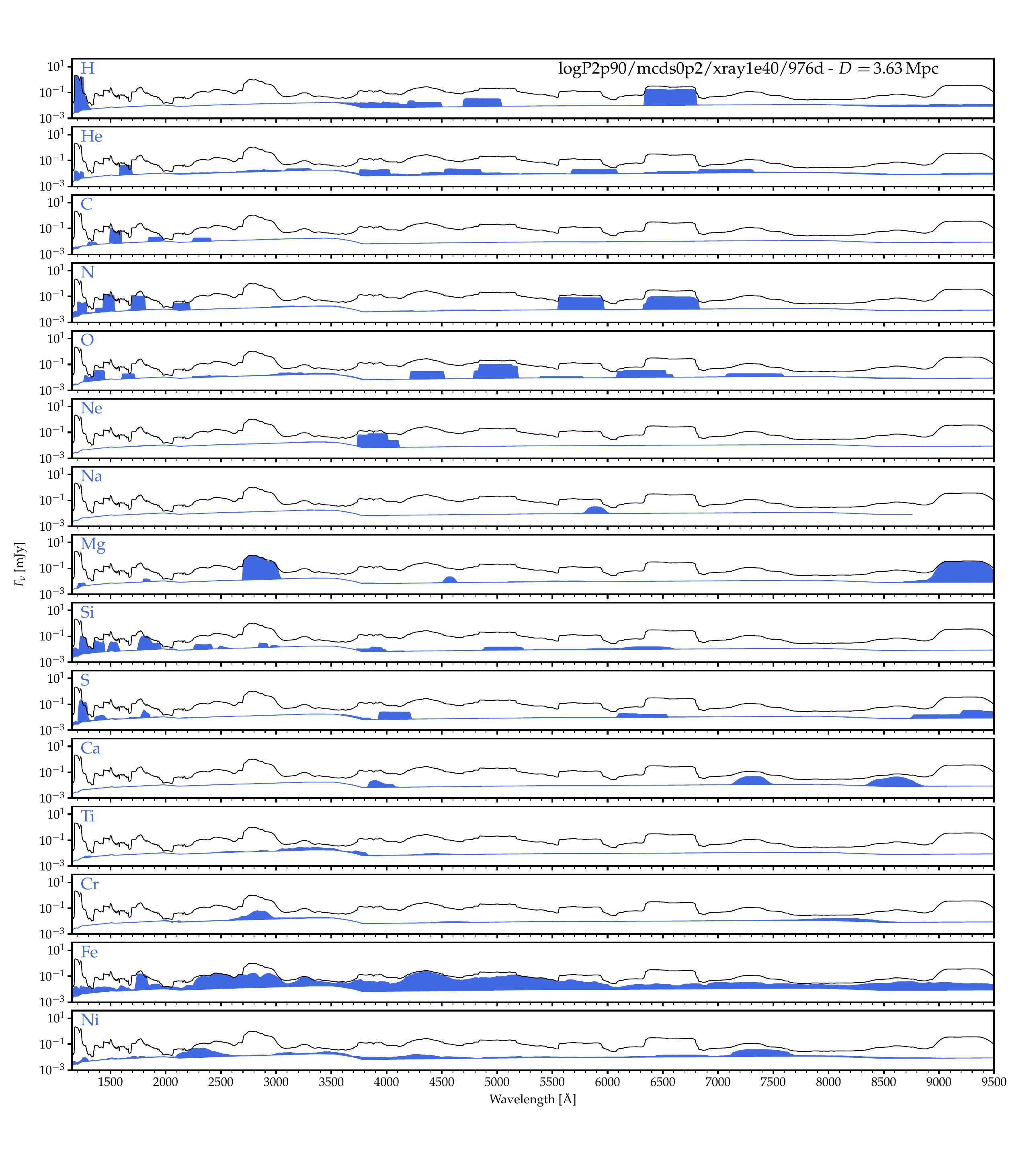}
\caption{Individual species contribution for model logP2p90/mcds0p2/xray1e40/976d  used for SN\,1993J and shown in Fig.~\ref{fig_93J}. Dust is ignored in the model shown.
\label{fig_93J_species}
}
\end{center}
\end{figure*}

\begin{figure*}
\begin{center}
\includegraphics[width=0.8\textwidth]{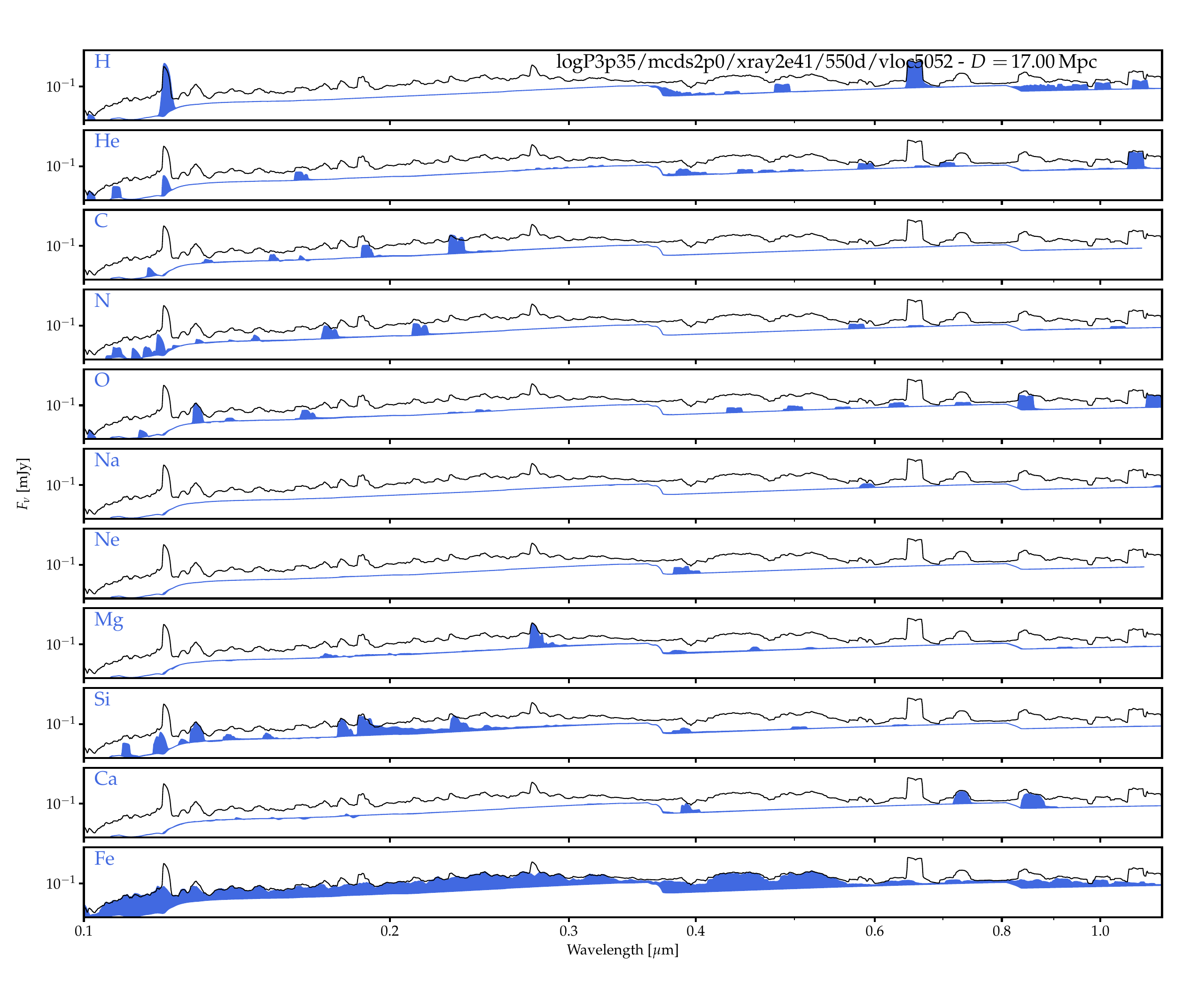}
\caption{Individual species contribution for (dust-free) model logP3p35/mcds2p0/xray2e41/550d/vloc5052  used for SN\,1998S and shown in Fig.~\ref{fig_98S}.
\label{fig_98S_species}
}
\end{center}
\end{figure*}

\begin{figure*}
\begin{center}
\includegraphics[width=0.8\textwidth]{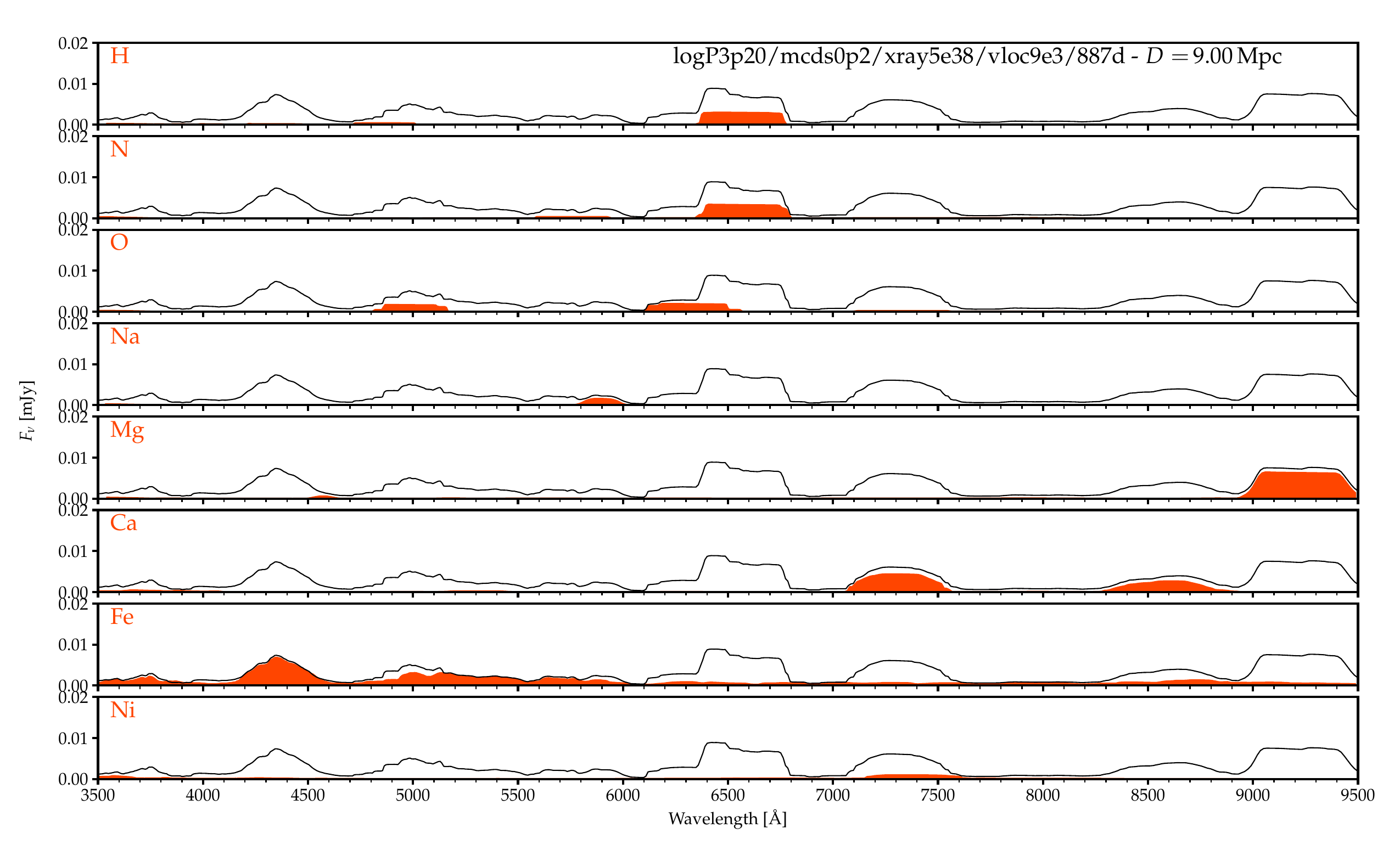}
\includegraphics[width=0.8\textwidth]{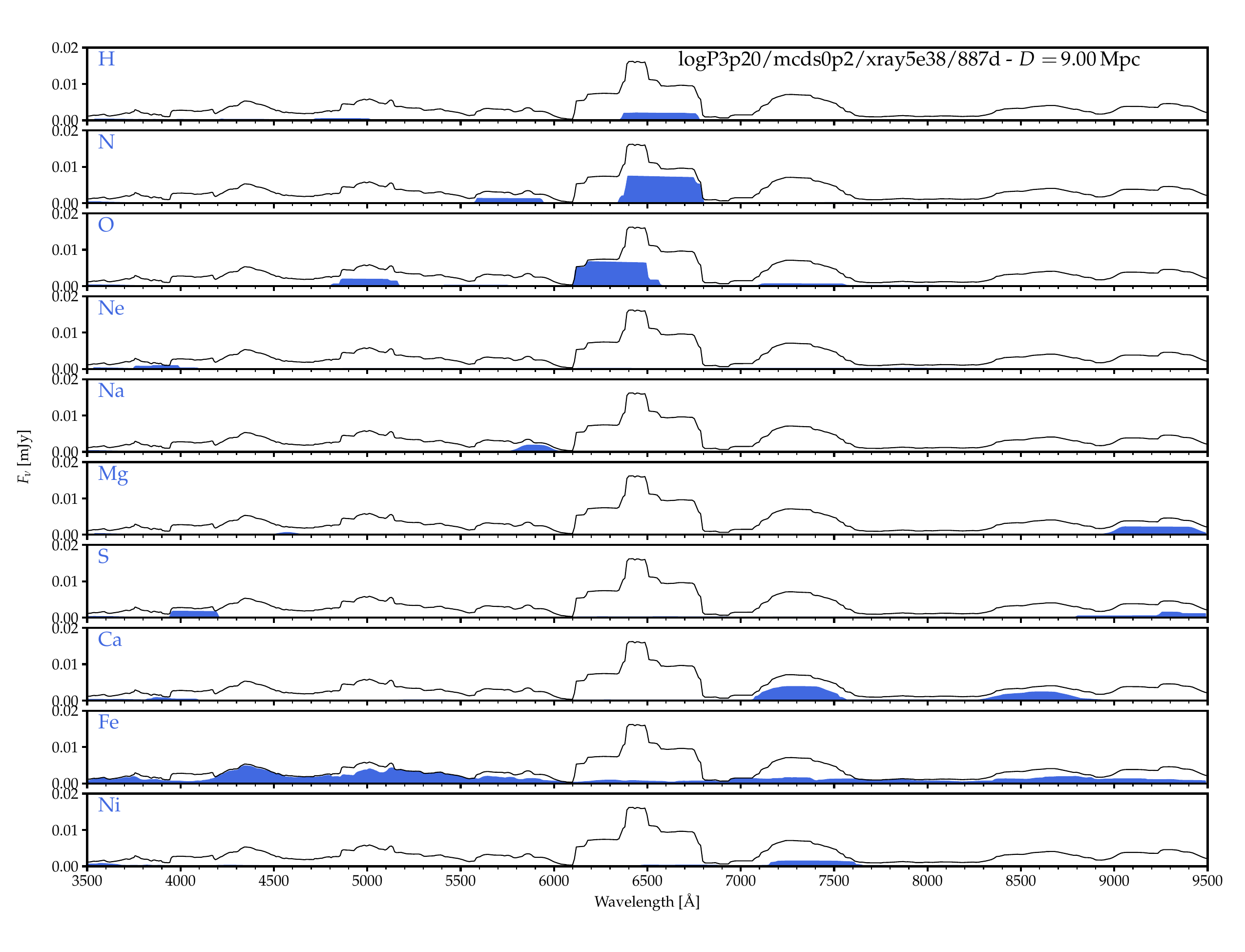}
\caption{Individual species contribution for model logP3p20/mcds0p2/xray5e38/vloc9e3 at 887\,d and model logP3p20/mcds0p2/xray5e38 used for SN\,2013ej and shown in Fig.~\ref{fig_13ej}.
\label{fig_13ej_species}
}
\end{center}
\end{figure*}

\begin{figure*}
\begin{center}
\includegraphics[width=0.8\textwidth]{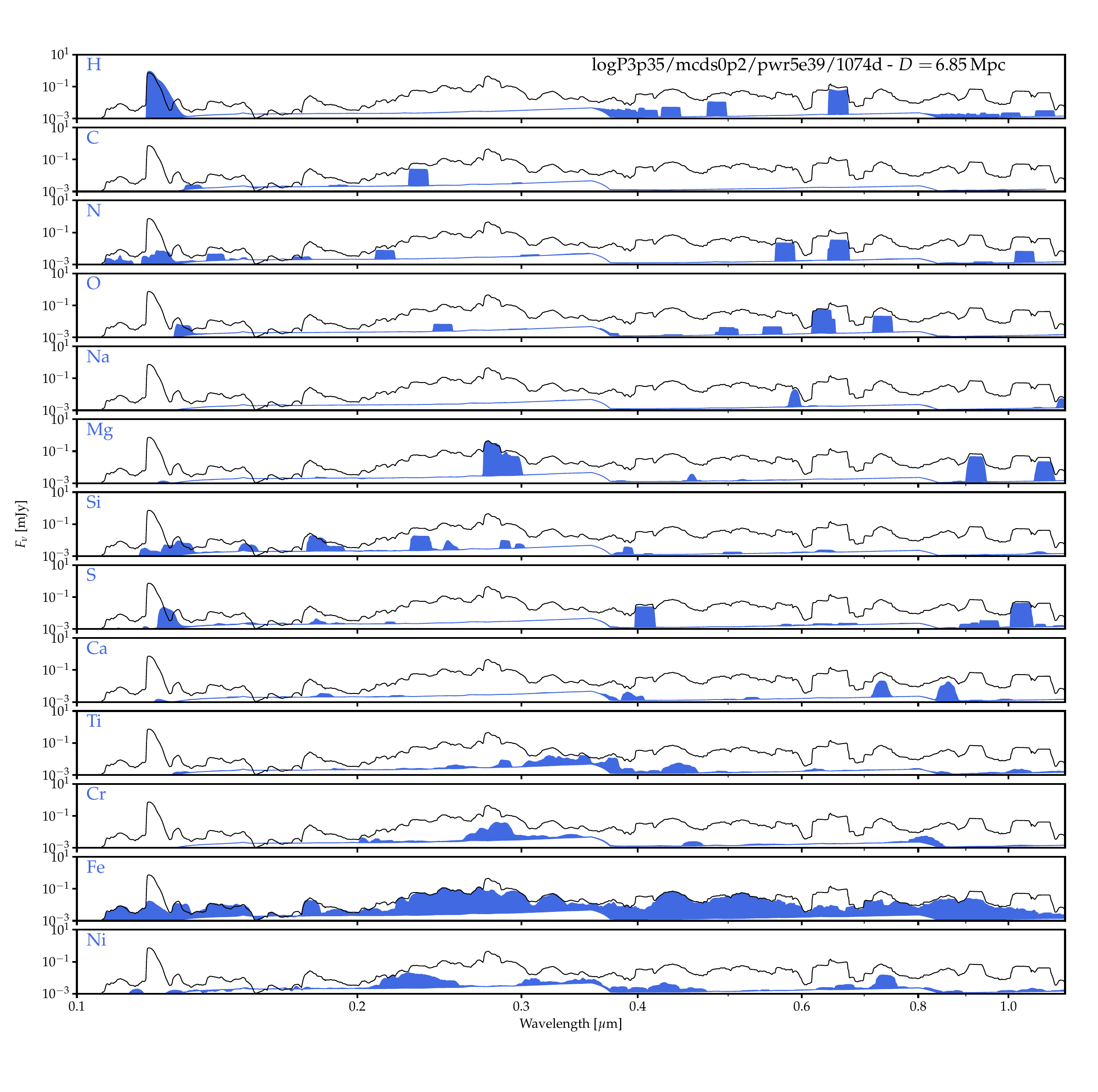}
\caption{Individual species contribution for model logP3p35/mcds0p2/pwr5e39 at 1074\,d used for SN\,2023ixf and shown in Fig.~\ref{fig_23ixf}.
\label{fig_23ixf_species}
}
\end{center}
\end{figure*}

\begin{figure*}
\begin{center}
\includegraphics[width=0.8\textwidth]{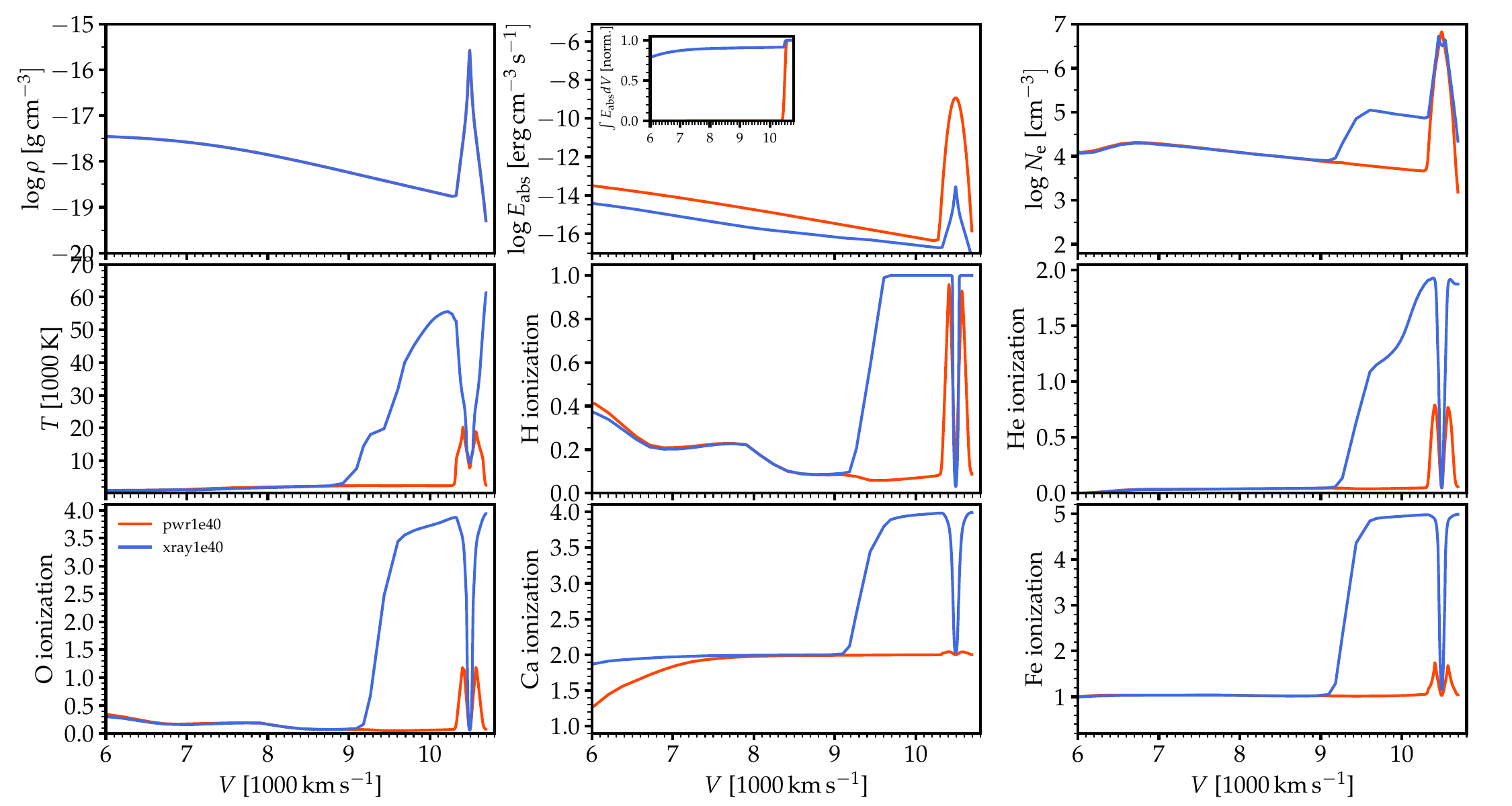}
\caption{Ejecta properties for the models shown in Fig.~\ref{fig_93J}. From the top row to bottom row and from left to right, we show the (clumped) mass density, the total power absorbed (this includes decay power and shock power deposited in the form of high energy electrons but not the X-ray power injected since this is stored in the radiative flux), the electron density, the gas temperature, followed by the ionization of H, He, O, Ca, and Fe (zero stands for the neutral state, one for once ionized etc.)
\label{fig_93J_prop}
}
\end{center}
\end{figure*}

\begin{figure*}
\begin{center}
\includegraphics[width=0.8\textwidth]{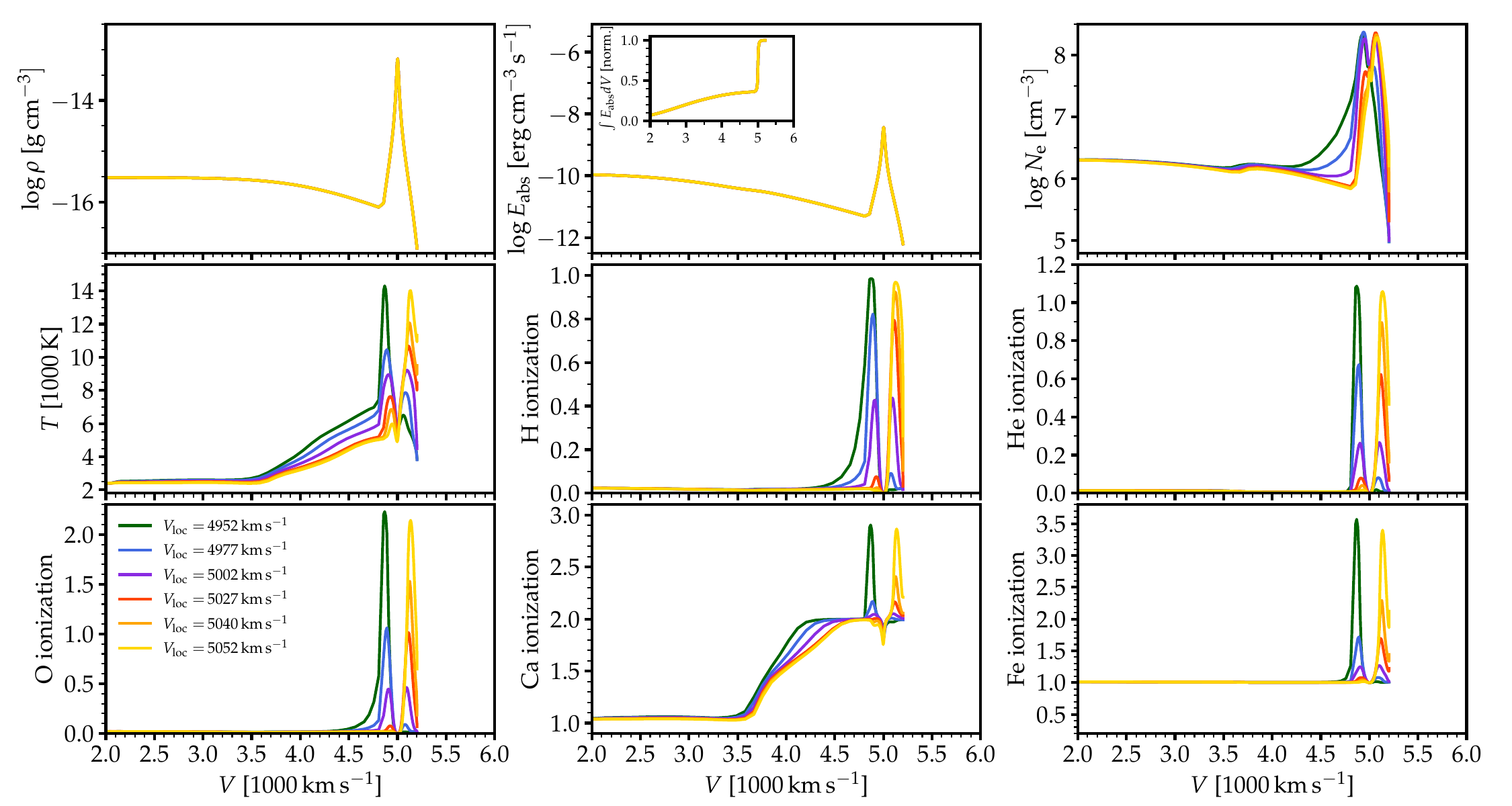}
\caption{Same as Fig.~\ref{fig_93J_prop} but for the models shown in Fig.~\ref{fig_98S_var_pwr}.
\label{fig_98S_prop}
}
\end{center}
\end{figure*}

\begin{figure*}
\begin{center}
\includegraphics[width=0.8\textwidth]{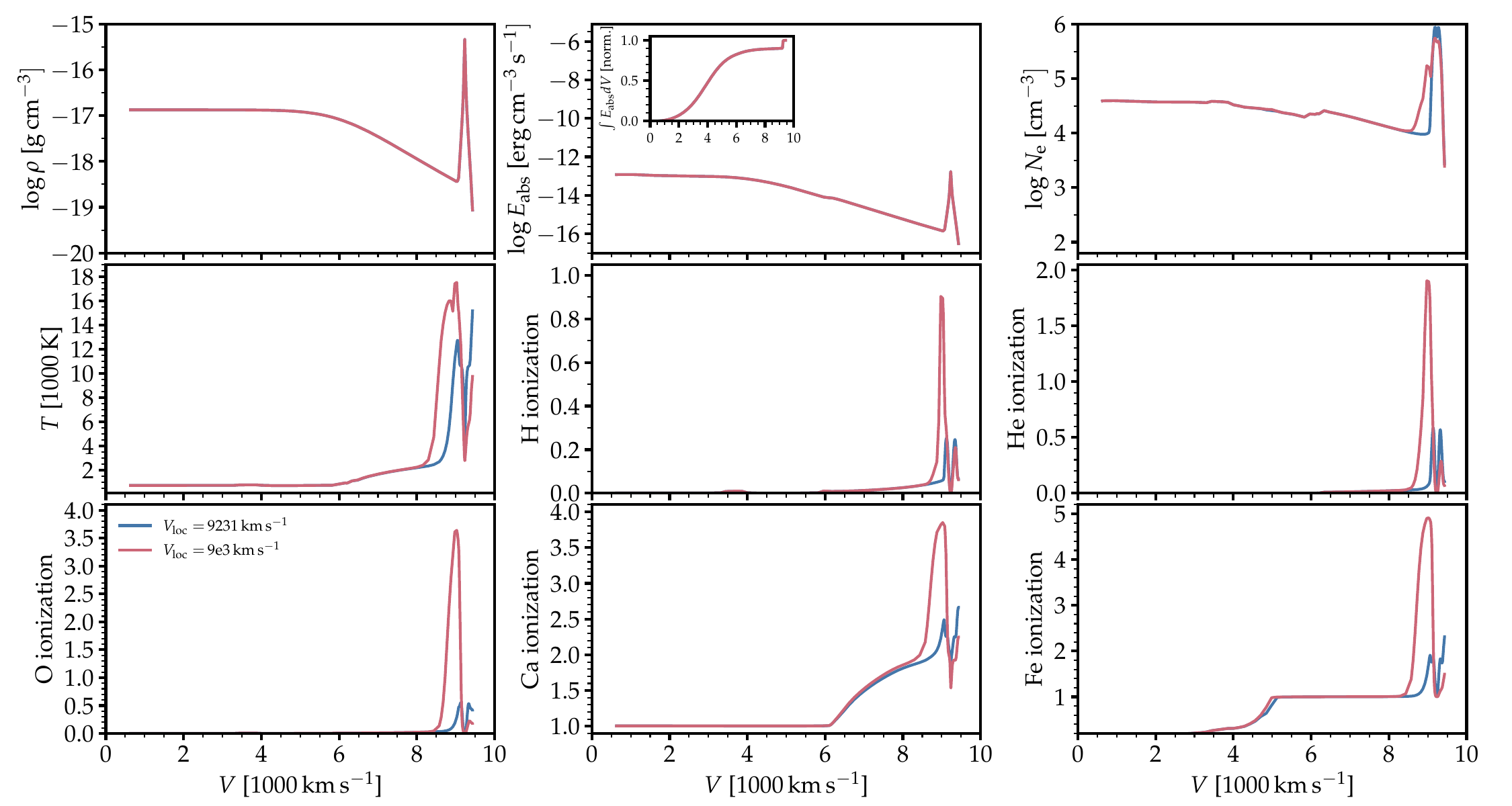}
\caption{Same as Fig.~\ref{fig_93J_prop} but for the two models shown in Fig.~\ref{fig_13ej}.
\label{fig_13ej_prop}
}
\end{center}
\end{figure*}

Figures~\ref{fig_93J_species}--\ref{fig_23ixf_species} show the individual species contributions in the best-fitting models to the observed SNe 1993J, 1998S, 2013ej, and 2023ixf discussed in Section~\ref{sect_obs}. In these illustrations, only the bound-bound transitions are included in the spectrum calculation, one species at a time. This is illustrative in the sense that the total flux is not the simple addition of all individual contributions. Figures~\ref{fig_93J_prop}--\ref{fig_13ej_prop} give some ejecta and CDS properties for these same models (except for the model used for SN\,2023ixf, whose modeling was described in detail in \citealt{dessart_23ixf_neb_26}). These figures allow one to assess the impact on the ionization (temperature, density of free electrons, species ionization) obtained under different conditions for the ejecta or CDS.

\end{document}